\documentclass[12pt]{article}

\usepackage{tikz}
\usepackage{epsfig,latexsym,amsfonts,amsmath,amsthm,amssymb,amsbsy,multirow,slashed,wasysym,textcomp,subfigure,wrapfig,datetime,comment,mathtools,cancel,cite,twistor,mathrsfs}
\usepackage[hidelinks]{hyperref}
\usetikzlibrary{calc}
\usepackage[font={footnotesize},bf]{caption}

\usepackage{float}

\usepackage[normalem]{ulem} 
\numberwithin{equation}{section}
\def\ee{\end{equation}}
\def\be{\begin{equation}}
\def\bea{\begin{eqnarray}}
\def\eea{\end{eqnarray}}
\newcommand{\beq}{\begin{eqnarray}}
\newcommand{\eqq}{\end{eqnarray}}
 \newcommand{\badat}{\begin{alignedat}}
 \newcommand{\eadat}{\end{alignedat}}

\newcommand{\eal}[1]{\be \begin{aligned} #1 \end{aligned}\end{equation}} 

\newcommand{\eqn}[1]{\be #1 \end{equation}} 
\newcommand{\eqa}[1]{\bea  #1\end{eqnarray}}

\long\def\new#1\endnew{{\bf #1}}		
\long\def\del#1\enddel{}

\def\del{\partial}

\let\polishl\l
\def\l{\lambda }

\def\tc{\theta_{\rm crit}}

\usepackage{color}

\newcommand{\pink}[1]{\textcolor{\pink}{#1}}

\definecolor{dblue}{rgb}{0.2,0.50,0.80}

\usepackage{xspace}

\def\sdym{SDYM}

\def\t2{T$^{1,1}$}

\catcode`,\active

\catcode`\,12

\usepackage{amsmath,amssymb}
\usetikzlibrary{arrows.meta,positioning,calc}
\usepackage[normalem]{ulem}

\tikzset{
  theory/.style={
    draw=black,
    line width=0.8pt,
    rounded corners=1.5pt,
    minimum height=2.35cm,
    text width=4.75cm,
    align=center,
    inner xsep=8pt,
    inner ysep=7pt,
    font=\normalsize
  },
  transition/.style={
    -{Stealth[length=3.2mm,width=2.2mm]},
    line width=0.9pt,
    black
  },
  equivalence/.style={
    {Stealth[length=3.2mm,width=2.2mm]}-{Stealth[length=3.2mm,width=2.2mm]},
    line width=0.9pt,
    black
  },
  transition label/.style={
    fill=white,
    inner sep=3pt,
    font=\normalsize,
    align=center
  }
}

\begin{document}
\begin{titlepage}
\unitlength = 1mm~\\
\vskip 3cm
\begin{center}


{\LARGE{Self-Dual Yang--Mills in AdS$_4$}}

\vspace{0.8cm}
Simon Heuveline$^{ab}$\footnote{simonheuveline@fas.harvard.edu}, Romain Ruzziconi$^{ab}$\footnote{romainruzziconi@fas.harvard.edu}, Ahmed Sheta$^{a}$\footnote{asheta@g.harvard.edu}, Andrew Strominger$^{abc}$\footnote{strominger@fas.harvard.edu}\\
\vspace{1cm}

{\it  $^a$Center for the Fundamental Laws of Nature, Harvard University, Cambridge, MA, USA\newline $^b$Black Hole Initiative, Harvard University, Cambridge, MA, USA\newline $^c$OpenAI, San Francisco, CA, USA}\\

\vspace{0.8cm}

\begin{abstract}

Quantum  Yang-Mills theory in Euclidean AdS$_4$ with Neumann boundary conditions is studied as a function of the complex couplings $ \frac{1}{g_\pm^2} \equiv \frac{1}{g^2}\mp \frac{i \theta }{8 \pi^2}$. We define  a ``self-dual limit"  by $g_- \to 0$ with $g_+ ~{\rm fixed}$. We argue  that the limiting  theory is well-defined perturbatively and equivalent to the BF formulation of self-dual Yang-Mills theory with a  particular boundary condition relating $B$ and $F$ at the boundary $\partial$AdS$_4$. $g_+$ is the loop-counting parameter of the self-dual theory. This is in stark contrast to flat space, where $\theta$ has no effect on perturbative dynamics. In the self-dual limit, it is shown  that all tree-level  zero-plus and single-plus boundary correlators vanish. A closed form expression is given for any number of gluons for the double-plus tree correlators. Finally, we show that in the flat space limit of AdS$_4$, the total energy poles in the tree boundary correlators correctly reduce to the  single-minus gluon amplitudes.

 \end{abstract}

\end{center}

\end{titlepage}

\setcounter{tocdepth}{2}
\tableofcontents

\section{Introduction}

Self-dual Yang--Mills theory (\sdym) and self-dual gravity in four
dimensions provide remarkably rich yet tractable subsectors of their
parent theories. Although genuinely nonlinear and interacting, they are
classically integrable and exactly solvable 
\cite{Penrose:1976js,Ward:1977ta,Ward:1985gz,MasonWoodhouse:1996}.
Much is also known about the perturbative quantum dynamics of these theories in flat space. At tree level, only the single-minus amplitudes are non-vanishing  \cite{Guevara:2026qzd,Guevara:2026qwa}.   Quantum corrections are given by a known one-loop exact, rational  expression \cite{Bardeen:1995gk,Chalmers:1996rq,Bern:1993sx}. 

Much less is known about the self-dual theories in AdS$_4$ or dS$_4$. Ward \cite{Ward:1985gz} showed that classical solutions in flat space have unique extensions to both, and a twistor description has been found  \cite{Ward:1980am,Alexandrov:2009vj,Adamo:2015ina}. 
 For self-dual
Yang--Mills, conformal invariance provides a direct local relation
between the equations on flat space and on conformally flat backgrounds
such as AdS$_4$ or dS$_4$ This was  exploited recently to map flat space soft algebras to light ray operator algebras in the CFT$_3$ dual of AdS$_4$ \cite{Sheta:2025oep}.
Holographic and cosmological correlators of spinning
fields in AdS$_4$ or dS$_4$, including correlators generated by
bulk Yang--Mills theory, have been studied in
\cite{Maldacena:2011nz,Raju:2011mp,Raju:2012zr,Raju:2012zs,
Albayrak:2018tam,Albayrak:2019asr,Baumann:2020dch,
Armstrong:2020woi,Albayrak:2020fyp,Baumann:2024ttn,Arundine:2026fbr,Gomez:2026yno}. Very interesting self-dual higher-spin theories in AdS$_4$ and 
their holographic observables have been recently studied in 
\cite{Lipstein:2023pih,Bittleston:2024rqe,Chowdhury:2024dcy,Skvortsov:2026gtq,Sharma:2025ntb}, including chiral limits similar or identical to those  considered here. References \cite{Aharony:2024nqs,Jain:2024bza} study  similar limits in an analytic continuation of the dual boundary CFT$_3$ in ABJM\cite{Aharony:2008ug}.

While the self-dual theories in AdS$_4$ are not well understood, one expects that there is a  quantum mechanically consistent generalization from flat space. Moreover, it is plausible that they have a holographic realization as a nonunitary CFT$_3$ living on the boundary of AdS$_4$. The construction of such a holographic dual would be extremely interesting.

As a first step, in this paper we consider  self-dual Yang-Mills theory in AdS$_4$, deferring the gravitational problem. Our first  goal is to understand something about the boundary correlators which one might hope to reproduce holographically. We encounter several interesting features.

 The first issue is that of AdS$_4$ boundary conditions, which generically break the SO(5,1) Euclidean conformal group down to SO(4,1).   These boundary conditions  become rather subtle at self-duality, so we invoke a limiting procedure. In this paper, we define SDYM theory in
AdS$_4$ through a limiting procedure applied to Yang--Mills theory with
a $\theta$ term, given by 
\begin{equation}\label{sdlintro}
    g_- \to 0 \;, \quad \text{at fixed $g_+$}\;,
\end{equation}
with 
\begin{equation} \label{def couplings}
    \frac{1}{g_\pm^2} \equiv \frac{1}{g^2}\mp \frac{i \theta }{8 \pi^2}~.
\end{equation}
This is equivalent to sending $g\to 0$ and $\theta$ to a critical value:
\begin{equation} \label{sdl2intro}
    \theta \to \theta_{\rm crit} =-i \frac{8 \pi^2}{g^2} \;, \quad g \to 0 \;, \quad  \text{with fixed } \frac{1}{g^2} - \frac{i \theta}{8 \pi^2} \;. 
\end{equation}
Sending $g_-\to 0$ freezes the anti-self dual fluctuations, while the self-dual ones are governed by $g_+$.

In section \ref{sec:Self-dual Yang-Mills in AdS} we show that Dirichlet boundary conditions are not deformed by the addition of a $\theta$ term. Dirichlet boundary conditions can never be compatible with self-duality: a positive helicity gluon reflects to a negative helicity one. Therefore, if we start at $\theta=0$ and deform to the self-dual couplings \eqref{sdlintro},  we can never reach a non-trivial self-dual theory. On the other hand, we show that, in the Neumann case, consistency of the variational principle requires deformation of the boundary conditions. At the limit, the boundary conditions imply the anti-self-dual part of the gauge field vanishes at the boundary, just as the doctor ordered. 

In order to understand the approach to the limit  in detail, we work out the Witten diagrams and compute the tree-level correlators through four points. Because $g_-$ is going to zero, some of the interaction vertices are going to infinity. Diagram by diagram, there are divergences, but we explicitly see them all cancel for tree correlators with up to four gluons in the limit \eqref{sdlintro}. At the same time, there is a dramatic simplification of the expressions for the correlators. 

In section \ref{sec:BF formulation}, we express the Witten rules for the Neumann case in the BF formalism for general $g_-$ and $g_+$. Importantly, there is a boundary condition relating $B$ and $F$ which breaks SO(5,1)  to SO(4,1). Unlike in the second-order formalism, here the divergences are absent diagram by diagram, and the self-dual limit is smooth. Hence one can study the theory directly at the self-dual point without invoking a limiting procedure. This is consistent with a formal path integral argument indicating equivalence between the two formalisms with carefulely chosen boundary conditions.  In principle, we need not have studied the second-order formalism at all, but because of the potential divergences, we sought  the direct corroboration.\footnote{ In principle there may be anomalies at the loop level, our formal path integral argument does not address these.} 

Non-perturbative effects are important in Yang-Mills theory, both from  confinement (for AdS$_4$ the radius must be small) and instantons. We do not study non-perturbative effects in this paper. However, due to the fact that the diagrammatic perturbation theory for quantum Yang-Mills theory agrees, for self-dual couplings and Neumann boundary conditions, with that of  BF self-dual Yang-Mills theory, we conjecture that 

{\it Quantum Yang-Mills with Neumann boundary conditions on AdS$_4$ analytically continued to  $g_-=0, ~~g_+ ~fixed$, is the same as  BF self-dual Yang-Mills on AdS$_4$ with coupling $g_+$. }

In this paper, we  provide a formal path integral argument for this conjecture.

Sections \ref{sec:Self-dual Yang-Mills in AdS} and \ref{sec:BF formulation} uncover  constraints on helicity configurations from diagrammatics.  At tree level, the single-plus and zero-plus amplitudes vanish for any number of gluons. Moreover,  a closed-form formula is given for the tree-level, color-ordered double-plus correlator at any multiplicity.  As discussed in \cite{Chowdhury:2024wwe, Baumann:2024ttn}, this compact arbitrary-multiplicity formula could play a key role in uncovering the structure of gluon correlators in AdS and dS. 

It is well known that, in the flat limit of AdS, the coefficients of the total energy pole in the boundary correlators are the flat space scattering amplitudes \cite{Maldacena:2011nz,Raju:2012zr,Raju:2012zs}. In section \ref{sec:Single-minus gluon amplitudes} we show that these pole coefficients are the single-minus amplitudes \cite{Guevara:2026qzd}, in  agreement with expectation. 

\begin{figure}[h!]
    \centering
\begin{tikzpicture}[node distance=2.2cm and 3.0cm]

  \node[theory] (secondorder) {
    Second-order Yang--Mills \\
    theory
    in $\mathrm{AdS}_4$ with a $\theta$ term \eqref{eq:action_theta} and Neumann boundary conditions \eqref{eq:N_F}
  };

  \node[theory, right=5.3cm of secondorder] (bf) {
    First-order BF formulation of Yang--Mills in $\mathrm{AdS}_4$ \eqref{eq:BF_action} with Neumann
    boundary conditions \eqref{eq:N_BF}
  };

  \coordinate (topmid) at ($(secondorder.east)!0.5!(bf.west)$);
  \draw[equivalence]
    (secondorder.east) --
    node[transition label, above=5pt] {Auxiliary field $B$ \eqref{EOMB}}
    (bf.west);

  \node[theory, below=3.05cm of topmid] (sdads) {
    SDYM\\
    in $\mathrm{AdS}_4$ with chiral\\
    boundary conditions
  };

   \draw[transition]
    (secondorder.south) --
    node[
      transition label,
      pos=0.7,
      above left=1pt and -2pt
    ]
    {Self-dual limit
     (Section~\ref{sec:Self-dual Yang-Mills in AdS})}
    (sdads.north west);

  \draw[transition]
    (bf.south) --
    node[
      transition label,
      pos=0.7,
      above right=1pt and -2pt
    ]
    {Self-dual limit
     (Section~\ref{sec:BF formulation})}
    (sdads.north east);

  \node[theory, below=2.45cm of sdads] (sdflat) {
    SDYM\\
    in flat spacetime
  };

  \draw[transition]
    (sdads.south) --
    node[transition label, right=5pt] {Flat-space limit (Section~\ref{sec:Single-minus gluon amplitudes})}
    (sdflat.north);

\end{tikzpicture}
\caption{SDYM in Euclidean AdS$_4$ is defined via a limiting procedure from first- and second-order formulations of Yang--Mills theory with Neumann boundary conditions. The flat-space limit of this theory reproduces the expected single minus amplitudes of SDYM in flat spacetime.}
    \label{fig:placeholder}
\end{figure}
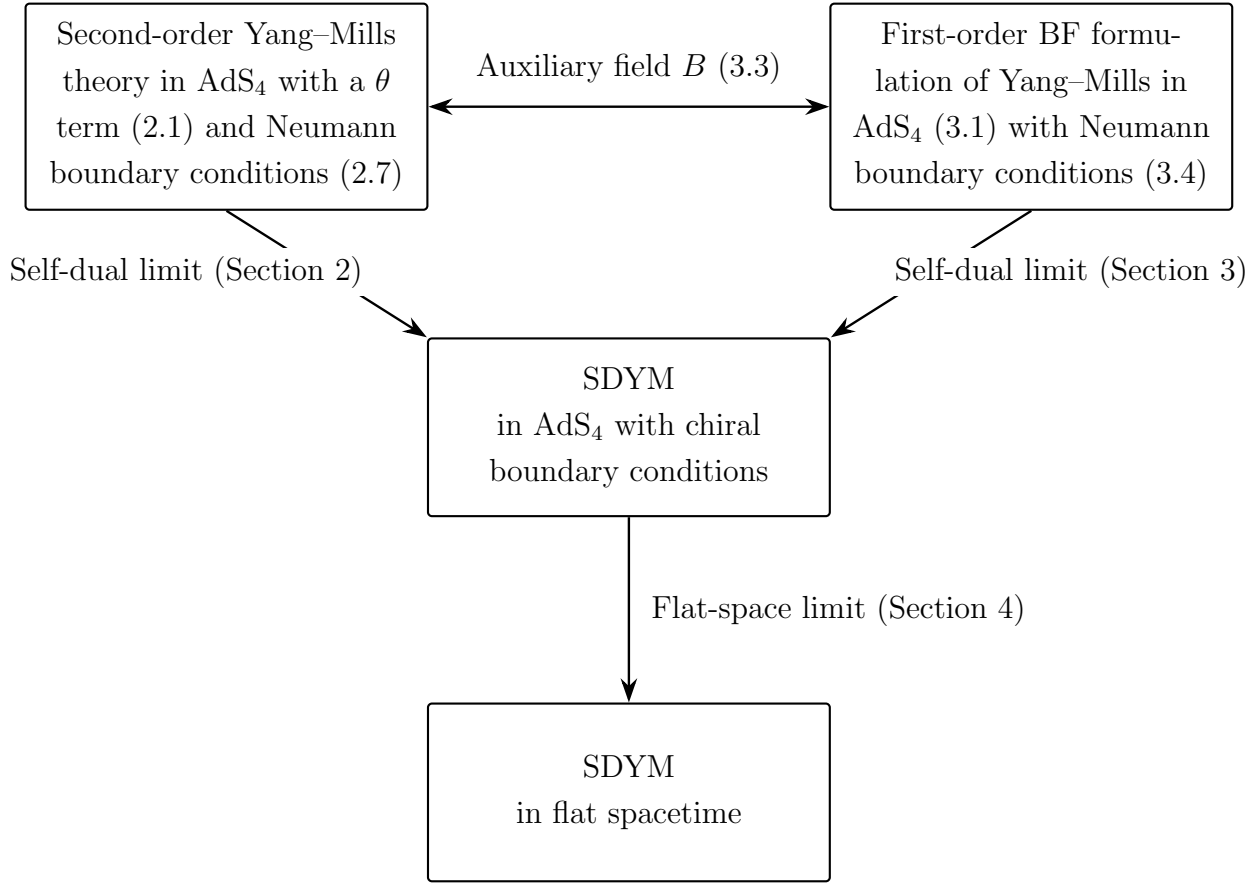

 The results of this paper are summarized in Figure~\ref{fig:placeholder}.


Our analysis mainly tackles the bulk perspective on the self-dual subsector of AdS$_4$ dynamics. Of course, as shown in \cite{Aharony:2024nqs,Jain:2024bza}, there should be a corresponding chiral subsector within the dual CFT$_3$, with simplifications in the conformal block expansion of the four-point function. This could be studied either through a bottom-up approach, assuming a CFT$_3$ built out of a current algebra, or in explicit top-down models such as ABJM \cite{Aharony:2008ug}.

\paragraph{Organization of the paper} In Section~\ref{sec:Self-dual Yang-Mills in AdS}, we explain features of Yang--Mills theory on AdS$_4$ that are absent in flat space and how they affect the self-dual limit, which we define by tuning $g$ and $\theta$ according to \eqref{sdl2intro}. We compute correlators through four points in this limit and show that they are finite. In Section~\ref{sec:BF formulation}, we explain the first-order BF formulation of Yang--Mills theory on AdS$_4$ and define the self-dual theory in this formulation, including its boundary conditions. We show that it is equivalent to the self-dual limit of the second-order formulation. We also derive a compact formula for the double-plus self-dual color-ordered tree-level correlator at arbitrary multiplicity. In Section~\ref{sec:Single-minus gluon amplitudes}, we show that the flat-space limit of our correlators reduces to the recently computed single-minus gluon amplitudes. The appendices contain computations omitted from the main text for brevity.


\vspace{0.2cm}

\emph{Note added:} While this paper was being prepared, interesting work by Skvortsov and Dongen \cite{Skvortsov:2026dru}, with significant overlap with the present work, appeared on arXiv.

\section{SDYM in AdS$_4$}
\label{sec:Self-dual Yang-Mills in AdS}

In this section, we  define\footnote{We do not know if  this definition is unique. } SDYM theory in AdS$_4$ by the analytic continuation  \eqref{sdlintro} or, equivalently, \eqref{sdl2intro}. 
This requires a careful treatment of the boundary conditions
and the associated boundary terms. We will show that the limit leads
to finite holographic correlators which drastically simplify and exhibit the expected chiral
properties of a self-dual theory.

\subsection{Boundary conditions}
\label{sec:Yang--Mills theory with}

In Euclidean signature, the Yang--Mills action with an arbitrary $\theta$ angle is given by
\begin{equation} \label{eq:action_theta}
    S [A] = \frac{1}{4 g^2} \int d^4 x \sqrt{g} \, F_{\mu \nu}^a F^{a,\mu\nu} -\frac{i \theta}{32 \pi^2} \int d^4 x \sqrt{g} \, F_{\mu \nu} ^a (\star F)^{a,\mu \nu}~,
\end{equation}
where $F_{\mu \nu}^a = \partial_\mu A_\nu^a-\partial_\nu A_\mu^a+  f^{abc} A_\mu^b A_\nu^c$. We can separate $F_{\mu \nu}^a$ into its self-dual and anti-self-dual components, 
\begin{equation}
    F^{\pm} = \frac{1}{2}(F \pm \star F)~.
\end{equation}
The action can then be written as 
\begin{equation} \label{eq:action_gpgm}
    S[A] = \frac{1}{4 g_+^2} \int d^4 x \sqrt{g} \,   F^{+,a}_{\mu \nu} F^{+,a; \mu \nu} + \frac{1}{4 g_-^2} \int d^4 x \sqrt{g} \, F^{-,a}_{\mu \nu} F^{-,a; \mu \nu}~,
\end{equation}
where, as stated above,
\begin{equation} \label{dico couplings}
   \frac{1}{g_\pm^2} \equiv \frac{1}{g^2}\mp \frac{i \theta }{8 \pi^2}~.
\end{equation}
Note that parity exchanges $g_+ \leftrightarrow g_-$.

Perturbative Yang--Mills amplitudes in flat space are independent of
$\theta$ because $F\wedge F$ is a total derivative which can be reduced to a boundary term.  The situation is quite different in AdS$_4$. Defining the AdS$_4$ theory requires a choice of boundary conditions, which affects the bulk physics already in perturbation theory. Consistent choices of boundary conditions depend on the value $\theta$, which thereby affects perturbative bulk correlators \cite{Maldacena:2011nz}.

The on-shell first variation of the action \eqref{eq:action_gpgm} is given by
\begin{equation} \label{eq:variation S}
    \delta S_{\text{on-shell}} = \frac{1}{g_+^2g_-^2} \int_{\partial (AdS_4)} d^3x \sqrt{\gamma} \, n_\mu  \delta A_\nu^a \left( g_-^2 F^{a,+,\mu \nu} + g_+^2 F^{a,-,\mu \nu} \right)~.
\end{equation}
There are two natural choices of boundary conditions that make the action stationary on shell. Dirichlet boundary conditions set
\begin{equation}\label{dir}
    \delta A_\nu^a \big|_{\partial (AdS_4)} = 0~,
\end{equation}
with $A_\nu^a \big|_{\partial (AdS_4)}$ taken to be pure gauge at the boundary. Alternatively, Neumann boundary conditions set
\begin{equation}
\label{eq:N_F}
    n_\mu \left(g_-^2 \, F^{a,+,\mu \nu} + g_+^2 \, F^{a,-,\mu \nu} \right) \bigg|_{\partial(AdS_4)} = 0~.
\end{equation}



When $\theta=0$ (or equivalently $g_+=g_-$), Dirichlet boundary conditions set ${n^\mu \star F_{\mu \nu}^a \big|_{\partial(AdS_4)} = 0}$, while the Neumann boundary conditions set ${n^\mu F_{\mu \nu}^a\big|_{\partial(AdS_4)}=0}$.\footnote{At the linearized level, the electric--magnetic duality of Maxwell theory acts as $SL(2,\mathbb{Z})$ on
$\tau\equiv \frac{\theta}{2\pi}+\frac{4\pi i}{g^2}$ while simultaneously
transforming the boundary conditions \cite{Witten:2003ya}. Pure
non-Abelian Yang--Mills theory does not retain this duality beyond the
linearized level, so different boundary conditions are not generically
equivalent. S-duality can be restored in special theories, most notably
$\mathcal N=4$ super Yang--Mills, but this lies beyond the scope of the
present work.} Turning on a $\theta$ angle does {\it not} affect the Dirichlet boundary conditions, which are consistent for any $\theta$, but shift the corresponding boundary current  $J^a_D \propto \delta S/\delta A^a$. By contrast, the $\theta$ angle modifies the Neumann boundary conditions as in \eqref{eq:N_F} but leaves the corresponding boundary operator unchanged. 

In the Dirichlet theory, the boundary value of $A$ is not integrated over, and one may freeze the color frame at the boundary. Boundary correlators of colored operators such as the current $J_D^a$ are then well defined.

With Neumann boundary conditions, however, the boundary operator is the gauge field $A^a_\mu\big|_{\partial (AdS_4)}$. From these basic gauge fields, one may construct gauge-invariant traced operators such as ${\rm Tr}F^2$ or Wilson loops. Correlators of the gauge field itself and its field strength, the central object of study herein, are {\it not} themselves gauge invariant. Nevertheless, they remain useful intermediate objects: they are  building blocks of gauge-invariant observables, flat-space gluon amplitudes can be extracted from their leading total-energy singularities (see Section~\ref{sec:Single-minus gluon amplitudes}) and they can be used to establish the perturbative equivalence of SDYM with a certain form of $B\wedge F$ theory. Moreover, their behavior as a function of $\theta$ provides a useful diagnostic of the nature of the self-dual limit, to which we now turn. 

\subsection{The self-dual limit }
\label{sec:Self-dual limit of Yang-Mills theory}

In this section, we define a self-dual limit of Yang--Mills theory in Euclidean AdS$_4$.
The self-duality condition $\star F = F$ is equivalent to
\begin{equation} \label{eq:SD}
    F^-=0~.
\end{equation}
In AdS, as discussed above, boundary conditions must be considered. 

Dirichlet boundary conditions \eqref{dir} are incompatible, for any  $\theta$, with nontrivial self-dual configurations:  imposing $n^\mu\star F_{\mu\nu}\big|_{\partial\mathrm{AdS}_4}=0$ along with the self-dual condition $F^-=0$ implies
$F|_{\partial\mathrm{AdS}_4}=0$. Regularity and uniqueness of the
first-order radial evolution then imply $F=0$ throughout the bulk.\footnote{ This can be explicitly seen from the propagator in \eqref{eq:J_Fpm_D} below.}

A similar argument conflicts self-duality with Neumann boundary conditions, {\it except} at the critical value $\theta=\tc$ where $g_-= 0$ and \eqref{eq:N_F}
reduces to 
\be {n^\mu F^- _{\mu \nu} \big|_{\partial\mathrm{AdS}_4}=0}~,\ee which is trivially compatible with the restriction to $F^-=0$.\footnote{In the self-dual theory with $F^-=0$ in the bulk, the condition $n^\mu F^-_{\mu \nu} \big|_{\partial}=0$ is vacuous. In Euclidean space, the equation of motion becomes first order and invertible, defining a unique Green's function that is regular in the interior.} The classical self-dual theory is defined by imposing \eqref{eq:SD} as
its equation of motion. By virtue of the Bianchi identity,
configurations satisfying \eqref{eq:SD} automatically obey the
second-order Yang--Mills equations.

Quantum Yang--Mills theory is formally defined by integrating
over all gauge-field configurations $A_\mu$, weighted by the action
\eqref{eq:action_gpgm}:
\begin{equation}
\label{second order path integral}
    Z_{\mathrm{YM}}
    =
    \int \mathcal D A \,
    \exp\left[
    -\frac{1}{4g_+^2}
    \int_{AdS_4} d^4x\,\sqrt{g}\,
    F^{+,a}_{\mu\nu}F^{+,a\,\mu\nu}
    -\frac{1}{4g_-^2}
    \int_{AdS_4} d^4x\,\sqrt{g}\,
    F^{-,a}_{\mu\nu}F^{-,a\,\mu\nu}
    \right]~,
\end{equation}
where the gauge-fixing and Faddeev--Popov factors are left implicit.\footnote{We will fix a gauge in the computations below, but will not need to keep track of the ghosts for  the tree correlators discussed here.}
We define quantum SDYM by a limiting procedure. Namely, we start from \eqref{second order path integral} and take the limit
\begin{equation} \label{eq:limit_g x}
     g_- \to 0,\qquad
     g_+~ \text{fixed }.
\end{equation}
 Formally, we have
\begin{equation}\label{lim}
\lim_{g_-\to0}
\mathcal N(g_-)
\exp\left[
-\frac{1}{4g_-^2}
\int_{AdS_4} d^4x\,\sqrt{g}\,
F^{-,a}_{\mu\nu}F^{-,a\,\mu\nu}
\right]
=
\delta[F^-]~,
\end{equation}
where $\mathcal N(g_-)$ is a field-independent normalization factor. Sending $g_-\to 0$ freezes anti-self-dual field configurations.
After removing this overall normalization, the path integral
\eqref{second order path integral} therefore reduces to
\begin{equation}
\label{sd path integral}
    Z_{\mathrm{YM}}^{\mathrm{sd}}
    =
    \int \mathcal D A\,\delta[F^-]\,
    \exp\left[
    -\frac{1}{4g_+^2}
    \int_{AdS_4} d^4x\,\sqrt{g}\,
    F^{+,a}_{\mu\nu} F^{+,a\,\mu\nu}
    \right]~.
\end{equation}
The path integral is supported on self-dual configurations
and provides a formal path-integral definition of self-dual
Yang--Mills theory. 

It is not manifestly obvious that the limit \eqref{lim} is well defined. Vertices contained in $(F^-)^2$ blow up as $g_-^{-2}$ and cause some diagrams to diverge. Cancellations among diagrams are required for a well-defined limit. Below, we verify explicitly that such cancellations occur at tree level. Moreover, we find that only certain helicity combinations can lead to nonzero limiting correlators.
In Section~\ref{sec:BF formulation}, we will show, at tree level, that the limiting theory defined by \eqref{sd path integral} is equivalent to the BF formulation of SDYM in AdS$_4$ with certain boundary conditions. This supports the notion that \eqref{sd path integral} is well-defined and corresponds to SDYM.

On the support of the delta functional, the action in \eqref{sd path integral} is topological and can be written as a boundary
Chern--Simons functional of level $ k_{\text{CS}}=\frac{4\pi i}{g_+^2}$ so that
\begin{equation}
    Z_{\mathrm{YM}}^{\mathrm{sd}}
    =
    \int \mathcal D A\,\delta[F^-]\,
    e^{-k_{\text{CS}} S_{\mathrm{CS}}[A]}~,
\end{equation}
where
\begin{equation}
    S_{\mathrm{CS}}[A]
    =
    \frac{i}{8\pi}
    \int_{z=0}d^3x\,\sqrt{\gamma} \,
    \epsilon^{ijk}
    \left(
        A_i^a\partial_j A_k^a
        +\frac{1}{3}f^{abc}A_i^aA_j^bA_k^c
    \right)~.
\end{equation}
Since $F^-=0$ is a first-order constraint, one may organize
the path integral by first integrating over bulk self-dual
configurations at fixed boundary value
$a_i=A_i|_{z=0}$ and then integrating over the boundary gauge field. This gives the formal expression:
\begin{equation}
    Z_{\mathrm{YM}}^{\mathrm{sd}}
    =
    \int\mathcal D a\,
    e^{-k_{\text{CS}} S_{\mathrm{CS}}[a]}\,
    \Psi_{\mathrm{SD}}[a],
    \qquad
    \Psi_{\mathrm{SD}}[a]
    \equiv
    \int_{A|_{\partial}=a}
    \mathcal D A\,\delta[F^-]~.
\label{eq:SD-boundary-wavefunctional}
\end{equation}
The wavefunctional is supported only on boundary data admitting a
self-dual extension. The Chern--Simons functional therefore
provides a topological boundary factor. The bulk wavefunctional
$\Psi_{\mathrm{SD}}[a]$ is generally nontrivial, and SDYM in AdS$_4$ is \emph{not} holographically dual to a pure Chern--Simons theory.

In particular, we could write the delta functional as a Lagrange multiplier path-integral
\begin{equation}
    \delta[F^-] = \int \mathcal D \tilde B \exp \left( {i\int d^4x \sqrt{g} \, \tilde B_{\mu \nu}^a F^{-,a~\mu \nu}} \right) \;,
\end{equation}
where $\tilde B$ is an anti-self-dual 2-form, such that\footnote{If we impose the gauge-fixing condition on $A$, we get the same contribution from Faddeev–Popov ghosts in the second-order path integral and the BF path integral.} 
\begin{equation} \label{eq:BF_act}
    \Psi_{\rm SD}[a] = \int_{A|_\partial=a} \mathcal DA \mathcal DB \; \exp  \left( - \int d^4 x \sqrt{g} \, B_{\mu \nu}^a F^{-,a~\mu \nu} \right) \;,
\end{equation}
where we rotated the contour $\tilde B =i B$. This is clearly a non-trivial functional of $a$, something we will verify explicitly in Section \ref{sec:BF formulation} by computing the diagrams of this theory. Performing the path integral over $a$ in $Z_{\rm YM}^{\rm sd}$ dynamically imposes the boundary condition
\begin{equation}
    n^\mu
    \left(
    B_{\mu\nu}^a
    +
    \frac{1}{2g_+^2}\star F_{\mu\nu}^a
    \right)
    \Big|_{\partial\mathrm{AdS}_4}
    =0~
\end{equation}
as the equation of motion of $a$ (which holds in correlators up to contact terms). We will discuss this BF formulation of the self-dual theory in AdS$_4$ in more detail in Section \ref{sec:BF formulation}.

Importantly, unlike its flat space counterpart, self-dual Yang-Mills in AdS$_4$ has a genuine coupling constant, $g_+^2$, which appears in the boundary CFT through the Chern-Simons level. This is evident from the path integral arguments above, and will show up in the correlators computed below. In fact, $g_+$ is the loop-counting parameter, which will be explained from the perspective of the BF formulation around \eqref{eq:loop_counting}.

For the rest of this section, we carry out the analysis in the second-order formulation, where we treat the self-dual theory as the $g_- \to 0$ limit of the full Yang-Mills theory defined as \eqref{second order path integral}.

\subsection{Propagators and vertices}
\label{sec:Propagators and vertices1}

This section presents the building blocks for the Witten diagrams of the full Yang--Mills theory in AdS$_4$ \eqref{eq:action_theta} with Neumann boundary conditions \eqref{eq:N_F} and any $\theta$.  

We work in Poincaré coordinates $x^\mu = (z,x^i)$ on Euclidean AdS$_4$, where the metric takes the form
\begin{equation} \label{eq:Poinc_metric}
    ds^2 = \frac{dz^2 + d\vec{x}^2}{z^2}~,
\end{equation}
and fix radial gauge:
\begin{equation}
    A_z = 0~. \label{radial gauge}
\end{equation}
Linear momentum is conserved in the $\vec x$, but not $z$, directions. 
The bulk-to-bulk propagator is obtained by inverting the kinetic term of Yang--Mills theory \eqref{eq:action_theta}. In $\vec x$ Fourier space at the boundary of AdS$_4$ \cite{Bzowski:2013sza,Bzowski:2019kwd} the propagator solves\footnote{Our conventions are $f(\vec{x}) = \int \frac{d^3 \vec{k}}{(2\pi)^3} \; e^{i \vec{k} \cdot \vec{x}} \tilde f(\vec{k})$, and we suppress an overall factor of $(2\pi)^3 \delta^3(\sum_i\vec{k}_i)$ in every momentum-space $n$-point function.}
\begin{equation}
    \left[(\partial_z^2 - k^2) \delta_{il} + k_i k_l  \right] \langle  A_l^a (z,  \vec k) A_j^b (z', - \vec k)    \rangle   = - g^2 \delta (z-z') \delta_{ij} \delta^{ab}~, \label{box in fourier}
\end{equation} where $\vec k$ denotes the three-dimensional Euclidean momentum and $k=|\vec k|$ its norm.  It is convenient to split the propagator as
\begin{equation}\label{dcmp}
    \langle  A_i^a (z,  \vec k) A_j^b (z', - \vec k)    \rangle = g^2 \delta^{ab} \left[ G^L (k, z, z') L_{ij} + G^+ (k, z, z') \Pi^+_{ij}  +   G^- (k, z, z') \Pi^-_{ij}  \right]~,
\end{equation} 
where
\begin{equation} \label{projectors}
     L_{ij} = \frac{k_i k_j}{k^2}, \qquad \Pi^\pm_{ij} = \frac{1}{2} \Big( \Pi_{ij} \pm h_{ij} \Big)  , \qquad \Pi_{ij} = \delta_{ij} - L_{ij} , \qquad  h_{ij} = \frac{i}{k} \epsilon_{ijm} k^m~, 
\end{equation}
where indices are lowered and raised by $\delta_{ij}$ and its inverse.
Here $L_{ij}$ projects onto the longitudinal subspace, $\Pi_{ij}$ projects onto the transverse subspace, $h_{ij}$ is the helicity operator, sometimes called the $\epsilon$-transform \cite{Caron-Huot:2021kjy,Jain:2021gwa}, and $\Pi^\pm_{ij}$ project onto its helicity eigenspaces, $h_i^{~k} \Pi^{\pm}_{kj} = \pm \Pi^{\pm}_{ij}$. Defining the polarization vectors $\epsilon^i_\pm(\vec{k})$ as
\begin{equation}
    k^i \epsilon_i^\pm(\vec{k}) = 0 \;, \quad \epsilon_{ijk} k_j \epsilon^\pm_k(\vec{k})=\mp i k \epsilon^\pm_i(\vec{k}) \;, \quad \epsilon^i_\pm(\vec{k})\epsilon_i^\mp(\vec{k}) = 1 \;,    
\end{equation}
we have
\begin{equation} \label{eq:polarization_vecs}
    k^i L_{ij} = k_j, \quad \epsilon^i_\pm L_{ij} = 0, \quad k^i \Pi^\pm_{ij} = 0 = \epsilon^i_\mp \Pi^\pm_{ij}, \quad \epsilon^i_\pm \Pi^\pm_{ij} = \epsilon^\pm_j~ \;.
\end{equation}

The decomposition \eqref{dcmp} yields the projected equations
\begin{equation} \label{eq:sources}
    \partial_z^2  G^L   = -  \delta (z-z') , \qquad (\partial_z^2 - k^2)  G^\pm   = - \delta (z-z')~. 
\end{equation}
The most general transverse solution compatible with regularity in the interior is given by
\begin{equation} \label{ansatz}
    G_\pm (k, z, z') = \frac{1}{2 k} (e^{-k |z'-z|} + r_{\pm}(k)  e^{-k |z'+ z|}  )~,
\end{equation} where $r_\pm (k)$ are functions to be fixed by boundary conditions. They correspond to inserting an image charge that is reflected across the AdS boundary $z' \to -z'$. This adds a source $\delta(z+z')$ to the right-hand side of \eqref{eq:sources}, which vanishes upon restricting to the AdS region $z>0$ where $e^{-k(z+z')}$ becomes a homogeneous solution. Similarly, the general longitudinal solution is
\begin{equation}
    G^L (k, z, z') = -\frac{1}{2} (|z'-z| + r_L (k) |z'+z| ) + l(k)~,
\end{equation} where $l(k)$ is an arbitrary function of $k$ reflecting the residual gauge ambiguity in radial gauge. 
The (linearized) Neumann boundary conditions \eqref{eq:N_F} imply
\begin{equation}
     r_\pm = 
     \left( \frac{g_+}{g_-} \right)^{\pm 2}, \qquad  r_L = 1~.
\end{equation} 
This corresponds to an image charge located at $-z'$.

Putting everything together, the bulk-to-bulk propagator reduces to 
\begin{align} \label{BB YM}
     \left\langle A_i^a(\vec k,z)A_j^b(-\vec k,z')\right\rangle &= \frac{g_+^2 g_-^2}{g_+^2+g_-^2} \delta^{ab} \frac{1}{k} \left[ \Pi^+_{ij} \left( e^{-k|z-z'|} + \frac{g_+^2}{g_-^2} e^{-k(z+z')} \right) \right. \\
         &\qquad\qquad\qquad\left. + \Pi^-_{ij} \left( e^{-k|z-z'|} + \frac{g_-^2}{g_+^2} e^{-k(z+z')} \right) + 2 k \,L_{ij} (l(k) - \max(z,z')) \right]~. \nonumber
\end{align}
Its transverse part can be written in a helicity basis,  $A_\pm^a(\vec{k}) \equiv \epsilon_\pm^i(\vec{k}) A_i^a(\vec{k})$, 
\begin{equation}
\begin{split}
     \left\langle A_\pm^a(\vec{k}, z) A^b_\pm (-\vec{k}, z') \right \rangle &=  \frac{g_+^2 g_-^2}{g_+^2+g_-^2} \delta^{ab} \frac{1}{k}  \left( e^{-k|z-z'|} + \frac{g_\pm^2}{g_\mp^2} e^{-k(z+z')} \right), \\
     \left\langle A_\pm^a(\vec{k}, z) A^b_\mp (-\vec{k}, z') \right \rangle &= 0~.
\end{split}
\end{equation}

Taking $z \to 0$ in \eqref{BB YM} leads to the bulk-to-boundary correlator
\begin{align} \label{Bb}
     \left\langle A_i^a(\vec k,0)A_j^b(-\vec k,z')\right\rangle &= \frac{g_+^2 g_-^2}{g_+^2+g_-^2} \delta^{ab} \frac{1}{k} \left[ \Pi^+_{ij} \left( e^{-k z'} + \frac{g_+^2}{g_-^2} e^{-kz'} \right) \right. \\
         &\qquad\qquad\qquad\left. + \Pi^-_{ij} \left( e^{-kz'} + \frac{g_-^2}{g_+^2} e^{-kz'} \right) + 2 k \,L_{ij} (l(k) - z') \right]~. \nonumber
\end{align}
The boundary operator associated with the Neumann boundary conditions is the boundary value of the gauge field $A_i^a (\vec k, 0)$, from which one can construct the three-dimensional field-strength tensor. Taking its three-dimensional Hodge dual, we define the boundary operator\footnote{Our conventions are $\epsilon_{zijk} \equiv \epsilon_{ijk}$.}
\begin{equation} \label{boundary current}
    \mathcal{J}^a_i (\vec k ) = -\frac{1}{2} \epsilon_{ijk} F_{jk}^a\Big|_{z=0} = - \star F_{zi}^{a}  \Big|_{z=0}~.
\end{equation} 
In the self-dual limit \eqref{eq:limit_g x}, we have $F^-|_{z=0} = 0$ and the above operator reduces to $\mathcal J^a_i (\vec k ) = - F_{zi}^{+, a} |_{z=0}$, which corresponds to self-dual boundary insertions. In perturbation theory, derivatives of  the propagator \eqref{Bb} are correlators of the linearized
version of \eqref{boundary current}, 
\begin{equation} \label{boundary current lin}
    J_i^a(\vec k)
    =
    -i\epsilon_{ijk}k_jA_k^a(\vec k,0)~,
\end{equation} which is conserved, $k_iJ_i^a(\vec k)=0$.\footnote{This operator is an exactly conserved $\Delta=2$ current in the Abelian theory.} This leads to the bulk-to-boundary propagator
\begin{equation} \label{Bb YM}
\left\langle
J_i^a(\vec k)A_j^b(-\vec k,z')
\right\rangle
=
\delta^{ab}e^{-kz'}
\left[
g_+^2\Pi^+_{ij}
-g_-^2\Pi^-_{ij}
\right]~,
\end{equation}
which is purely transverse. Bulk self-duality is correlated with boundary helicity:\footnote{Had we used Dirichlet boundary conditions instead, the bulk-to-boundary propagator would have been
\begin{equation} \label{eq:J_Fpm_D}
   \left\langle
J_{D,i}^a(\vec k)F_j^{\pm,b}(-\vec k,z')
\right\rangle_D \propto \delta^{ab} g^2 \, k e^{-kz'} \Pi^\pm_{ij}~,
\end{equation}
which shows that with Dirichlet boundary conditions, $F^-$ could not be frozen to zero in the bulk while allowing a fluctuating $F^+$.}

\begin{equation} \label{eq:J_Fpm_N}
    \left\langle
J_i^a(\vec k)F_{zj}^{\pm,b}(-\vec k,z')
\right\rangle = \mp \delta^{ab}  g_\pm^2 \, k\,  e^{-k z'} \Pi^{\pm}_{ij}~.
\end{equation}

Yang--Mills theory with a $\theta$ term \eqref{eq:action_theta} has three vertices. The pure Yang--Mills term yields a cubic vertex\footnote{Again, we suppress an overall factor of $(2\pi)^3 \delta^3(\sum_i \vec{k}_i)$ in every vertex.}
\begin{equation}
\begin{split}
\mathcal{V}^{abc,\mathrm{YM}}_{ijk}
(\vec{k}_1,\vec{k}_2,\vec{k}_3)
={}& 
-\frac{i}{2}
\left(
\frac{1}{g_+^2}+\frac{1}{g_-^2}
\right)
f^{abc}
\Big[
\delta_{ij}(k_1-k_2)_k
\\
&\qquad
+\delta_{jk}(k_2-k_3)_i
+\delta_{ki}(k_3-k_1)_j
\Big] \int_0^\infty dz~,
\end{split} \label{cubic}
\end{equation} as well as a quartic vertex
\begin{equation}
\label{eq:quarticVert}
\begin{split}
\mathcal{V}^{abcd,\mathrm{YM}}_{ijkl}
={}& 
-
\frac{1}{2}
\left(
\frac{1}{g_+^2}+\frac{1}{g_-^2}
\right)
\Big[
f^{abe}f^{cde}
\left(
\delta_{ik}\delta_{jl}
-\delta_{il}\delta_{jk}
\right)
\\
&\qquad
+f^{ace}f^{bde}
\left(
\delta_{ij}\delta_{kl}
-\delta_{il}\delta_{jk}
\right)
\\
&\qquad
+f^{ade}f^{bce}
\left(
\delta_{ij}\delta_{kl}
-\delta_{ik}\delta_{jl}
\right)
\Big] \int_0^\infty dz~.
\end{split}
\end{equation} 
For future convenience, we also include the $\mathrm{tr}(T^aT^bT^cT^d)$ color-ordered quartic vertex, which can easily be read off from \eqref{eq:quarticVert}:
\begin{equation}
\begin{aligned}
\mathcal{V}^{\mathrm{YM}}_{ijkl}
&={} 
-\frac{1}{2}
\left(
\frac{1}{g_+^2}+\frac{1}{g_-^2}
\right)
\left(
2\delta_{ik}\delta_{jl}-\delta_{il}\delta_{jk}- \delta_{ij}\delta_{kl}
\right)\int_0^\infty dz \\
&= -\frac{1}{2} \left(
\frac{1}{g_+^2}+\frac{1}{g_-^2}
\right) (\epsilon_{ijm} \epsilon_{klm}+ \epsilon_{lim} \epsilon_{jkm}) \int _0^\infty dz~.
\end{aligned} \label{quartic}
\end{equation} 
The $\theta$ term, which can be rewritten as a Chern--Simons boundary term, yields one cubic vertex evaluated at $z=0$: 
\begin{equation} \label{cubic CS}
\mathcal{V}^{abc,\mathrm{CS}}_{ijk}
=
\frac{1}{2}
\left(
\frac{1}{g_+^2}-\frac{1}{g_-^2}
\right)
f^{abc}\epsilon_{ijk}~.
\end{equation}
There is no quartic vertex in the $F\wedge F$ term because of the Jacobi identity $f^{abe}f^{cde} A^a\wedge A^b\wedge A^c\wedge A^d =0$, consistent with the fact that the $F\wedge F$ term reduces to a boundary Chern--Simons term. The vertices are decomposed into a helicity basis in Appendix \ref{sec:Cancellation of divergences in the self-dual limit}.

\begin{figure}[h!]
    \centering
\includegraphics[width=1.0\textwidth]{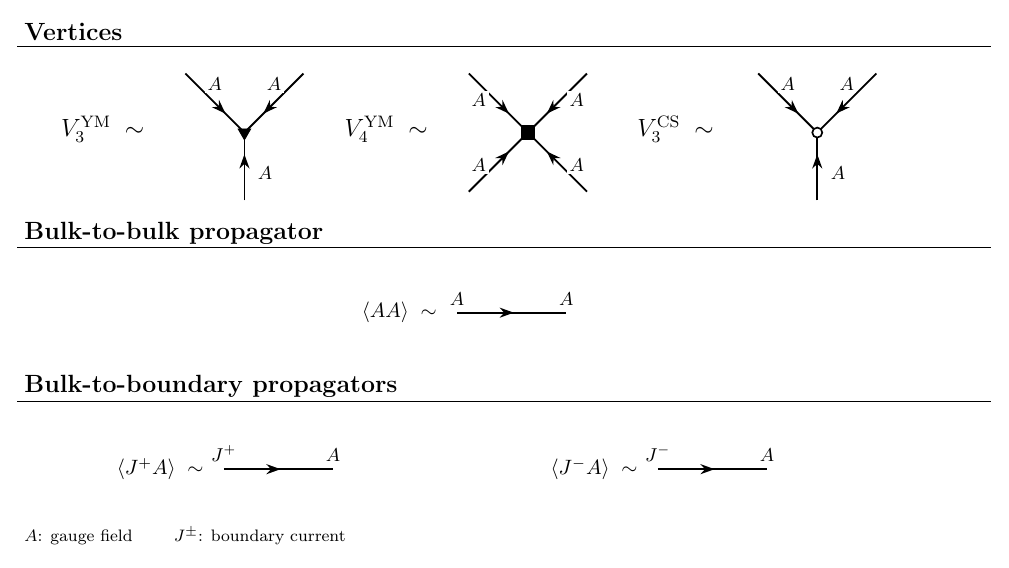}
\caption{Rules for Witten diagrams associated with the Yang--Mills theory \eqref{eq:action_theta} at finite values of the couplings.}
\label{fig:Feynman2nd}
\end{figure}

The building blocks of the Witten diagrams associated with the second-order Yang--Mills theory with a $\theta$ term are displayed in Figure~\ref{fig:Feynman2nd}.

Notice that, although the bulk-to-bulk propagators \eqref{BB YM} and
the bulk-to-boundary propagators \eqref{Bb YM} remain finite in the
self-dual limit \eqref{eq:limit_g x}, each of the three interaction
vertices diverges as $g_-^{-2}$. Hence there are no Feynman rules available {\it at} $g_-=0$.  Moreover it is  not manifest,
diagram by diagram, that the self-dual limit of Yang--Mills theory yields finite correlators. In the next
section, we show that the divergent contributions cancel up to four points once all
diagrams are combined, provided that the subleading $g_-$-dependent
terms in the propagators are retained. The resulting finite
correlators simplify drastically in the self-dual limit. This provides evidence for our definition of the self-dual theory as the $g_- \to 0$ limit of Yang-Mills, which will be further corroborated in Section \ref{sec:BF formulation} by comparing to the BF formulation.

\subsection{Boundary correlators in SDYM}
\label{sec:Self-dual holographic correlators}

In this subsection, we compute the two-, three-, and four-point tree-level boundary correlators of the operator \eqref{boundary current lin} in SDYM. The corresponding expressions for the full non-linear operator \eqref{boundary current} will be discussed in Section \ref{sec:CorrelatorsofJ}. As we will see, although the non-linear corrections contribute at tree level, they do not qualitatively modify the results presented here.

\paragraph{Two-point function.} The two-point function can be obtained from the bulk-to-boundary propagator \eqref{Bb YM} by taking $z' \to 0$ and using the definition of the linearized boundary current \eqref{boundary current lin}.
We find
\begin{equation}
\label{eq:JJ-full-YM}
\left\langle
J_i^a(\vec k)J_j^b(-\vec k)
\right\rangle
=
\delta^{ab}k
\left[
g_+^2\Pi^+_{ij}(\vec k)
+g_-^2\Pi^-_{ij}(\vec k)
\right]~.
\end{equation}
This expression is transverse, $k_i
\left\langle
J_i^a(\vec k)J_j^b(-\vec k)
\right\rangle
=0$. It is instructive to decompose this two-point function into helicity eigenstates \eqref{eq:polarization_vecs}.
Contracting the boundary operators in \eqref{eq:JJ-full-YM} with $\epsilon^i_\pm$, 
\begin{equation}
    J_\pm^a(\vec{k}) \equiv \epsilon^i_\pm(\vec{k}) J_i^a(\vec{k})~,
\end{equation}
we find 
\begin{equation}
\begin{split}
    \langle J^a_+ (\vec k) J^b_+ (- \vec k)   \rangle = \delta^{ab} g_+^2 \,k , \qquad  \langle J^a_+ (\vec k) J^b_- (- \vec k)   \rangle = 0, \qquad \langle J^a_- (\vec k) J^b_- (- \vec k)   \rangle = \delta^{ab} g_-^2 \,k~.
\end{split}
\end{equation} 
Finally, taking the self-dual limit \eqref{eq:limit_g x}, the two-point function simplifies and becomes chiral
\begin{equation} \label{2ptfunction}
\begin{split} 
    \langle J^a_+ (\vec k) J^b_+ (- \vec k)   \rangle = \delta^{ab} g_+^2\,k , \qquad  \langle J^a_+ (\vec k) J^b_- (- \vec k)   \rangle = 0, \qquad \langle J^a_- (\vec k) J^b_- (- \vec k)   \rangle = 0~.
\end{split}
\end{equation}
This result agrees with previous discussions of self-dual theories in AdS$_4$ \cite{Jain:2024bza,Aharony:2024nqs} and is a first manifestation of the simplification arising in the self-dual limit.

\paragraph{Three-point function.} In the following, we focus on color-ordered correlators and strip off the color factors. There are two contributing diagrams at three points: the bulk Yang--Mills cubic contact diagram and the boundary Chern--Simons contact diagram. At fixed $g_+$, the external bulk-to-boundary
propagators behave as
\begin{equation}
    \langle J^+A\rangle
    \sim \mathcal O(1),
    \qquad
    \langle J^-A\rangle
    \sim \mathcal O(g_-^2)~,
\end{equation}
while the interaction vertices satisfy
\begin{equation}
    \mathcal V_{\mathrm{YM}}^{(3)}
    \sim
    \mathcal V_{\mathrm{CS}}^{(3)}
    \sim
    \mathcal V_{\mathrm{YM}}^{(4)}
    \sim
    \mathcal O(g_-^{-2})~.
\end{equation} Hence the only potential divergence in the self-dual limit \eqref{eq:limit_g x} comes from the $\langle J^+ J^+ J^+ \rangle$ sector. We have explicitly:
\begin{align}
\left\langle
J^+(\vec k_1)J^+(\vec k_2)J^+(\vec k_3)
\right\rangle_{\mathrm{YM}}
&=
\frac{g_+^6}{2}
\left(
\frac{1}{g_+^2}
+
\frac{1}{g_-^2}
\right)
\epsilon_{ijk}\,
\epsilon_i^+(\vec k_1)
\epsilon_j^+(\vec k_2)
\epsilon_k^+(\vec k_3),
\\
\left\langle
J^+(\vec k_1)J^+(\vec k_2)J^+(\vec k_3)
\right\rangle_{\mathrm{CS}}
&=
\frac{g_+^6}{2}
\left(
\frac{1}{g_+^2}
-
\frac{1}{g_-^2}
\right)
\epsilon_{ijk}\,
\epsilon_i^+(\vec k_1)
\epsilon_j^+(\vec k_2)
\epsilon_k^+(\vec k_3)~.
\end{align}
Each of these diagrams diverges in the self-dual limit \eqref{eq:limit_g x}, but their sum is finite. After summing all diagrams, the different helicity configurations are
\begin{equation} 
    \begin{split}
\left\langle
J^+(\vec k_1)J^+(\vec k_2)J^+(\vec k_3)
\right\rangle
&=
g_+^4\,
\epsilon_{ijk}\,
\epsilon_i^+(\vec k_1)
\epsilon_j^+(\vec k_2)
\epsilon_k^+(\vec k_3),
\\[0.5em]
\left\langle
J^-(\vec k_1)J^+(\vec k_2)J^+(\vec k_3)
\right\rangle
&=g_+^2
\frac{
g_+^2k_1-g_-^2(k_2+k_3)
}{
k_1+k_2+k_3
}
\epsilon_{ijk}\,
\epsilon_i^-(\vec k_1)
\epsilon_j^+(\vec k_2)
\epsilon_k^+(\vec k_3),
\\[0.5em]
\left\langle
J^-(\vec k_1)J^-(\vec k_2)J^+(\vec k_3)
\right\rangle
&=g_-^2
\frac{
g_-^2k_3-g_+^2(k_1+k_2)
}{
k_1+k_2+k_3
}
\epsilon_{ijk}\,
\epsilon_i^-(\vec k_1)
\epsilon_j^-(\vec k_2)
\epsilon_k^+(\vec k_3),
\\[0.5em]
\left\langle
J^-(\vec k_1)J^-(\vec k_2)J^-(\vec k_3)
\right\rangle
&=
g_-^4\,
\epsilon_{ijk}\,
\epsilon_i^-(\vec k_1)
\epsilon_j^-(\vec k_2)
\epsilon_k^-(\vec k_3)~.
\end{split}
\end{equation}
Taking the self-dual limit \eqref{eq:limit_g x}, these correlators simplify drastically:
\begin{equation} \label{3pt correlators}
    \begin{split}
\left\langle
J^+(\vec k_1)J^+(\vec k_2)J^+(\vec k_3)
\right\rangle
&=
g_+^4\,
\epsilon_{ijk}\,
\epsilon_i^+(\vec k_1)
\epsilon_j^+(\vec k_2)
\epsilon_k^+(\vec k_3),
\\[0.5em]
\left\langle
J^-(\vec k_1)J^+(\vec k_2)J^+(\vec k_3)
\right\rangle
&=g_+^4
\frac{
k_1}{
E
}
\epsilon_{ijk}\,
\epsilon_i^-(\vec k_1)
\epsilon_j^+(\vec k_2)
\epsilon_k^+(\vec k_3),
\\[0.5em]
\left\langle
J^-(\vec k_1)J^-(\vec k_2)J^+(\vec k_3)
\right\rangle
&=0 ,
\\[0.5em]
\left\langle
J^-(\vec k_1)J^-(\vec k_2)J^-(\vec k_3)
\right\rangle
&=0~,
\end{split}
\end{equation} where $E = k_1 + k_2+ k_3$.

\paragraph{Four-point function.} The four-point correlator is of special interest because it is the first one which is not fixed by the boundary SO(4,1) conformal invariance: it involves bulk-to-bulk exchanges and therefore probes more deeply the nature of the bulk theory. The cubic Yang--Mills and Chern--Simons vertices enter the exchange diagrams, while the quartic Yang--Mills vertex enters the contact diagram. Since we consider color-ordered boundary correlation functions, only the $s$- and $t$-channels contribute. In our conventions, the $t$-channel contribution is obtained by exchanging external legs $2$ and $4$ in the $s$-channel contribution. In what follows, we consider color-stripped boundary correlation functions.

The leading $\Pi^+$ image contribution to the internal $AA$
propagator is of order $\mathcal O(1)$. By contrast, the subleading
$\Pi^+$ direct contribution, the $\Pi^-$ contribution, and the longitudinal
part of the propagator start at order $\mathcal O(g_-^2)$. It follows that an exchange diagram containing two cubic vertices and
$N_-$ external negative-helicity currents generically behaves as
\begin{equation}
    \left.
    \langle J^4\rangle
    \right|_{\mathrm{exchange}}
    \sim
    \mathcal O\left(g_-^{2N_--4}\right)~,
\end{equation}
whereas the quartic Yang--Mills contact diagram behaves as
\begin{equation}
    \left.
    \langle J^4\rangle
    \right|_{\mathrm{contact}}
    \sim
    \mathcal O\left(g_-^{2N_--2}\right)~.
\end{equation}
The contributions from the subleading parts of the internal
propagator are
\begin{equation}
    \left.
    \langle J^4\rangle
    \right|_{\mathrm{exchange,\,subleading}}
    \sim
    \mathcal O\left(g_-^{2N_--2}\right)~.
\end{equation}
The behavior of the individual diagrams is summarized in
Table~\ref{tab:gminus-scaling}. Here, $\mathrm{YM}_4$ denotes the
quartic Yang--Mills contact diagram, while
$\mathrm{YM}\!-\!\mathrm{CS}$ and
$\mathrm{CS}\!-\!\mathrm{YM}$ distinguish which cubic vertex appears
on the left and right sides of the exchange diagram. The same power
counting applies separately to the $s$- and $t$-channel diagrams.

\begin{table}[t]
    \centering
    \small
    \setlength{\tabcolsep}{5pt}
    \renewcommand{\arraystretch}{1.3}
    \begin{tabular}{c|ccccc}
        \hline
        Helicity
        &
        $\mathrm{YM}\!-\!\mathrm{YM}$
        &
        $\mathrm{YM}\!-\!\mathrm{CS}$
        &
        $\mathrm{CS}\!-\!\mathrm{YM}$
        &
        $\mathrm{CS}\!-\!\mathrm{CS}$
        &
        $\mathrm{YM}_4$
        \\
        \hline
        $++++$
        &
        $\mathcal O(g_-^{-4})$
        &
        $\mathcal O(g_-^{-4})$
        &
        $\mathcal O(g_-^{-4})$
        &
        $\mathcal O(g_-^{-4})$
        &
        $\mathcal O(g_-^{-2})$
        \\
        $-+++$
        &
        $\mathcal O(g_-^{-2})$
        &
        $\mathcal O(g_-^{-2})$
        &
        $\mathcal O(g_-^{-2})$
        &
        $\mathcal O(g_-^{-2})$
        &
        $\mathcal O(g_-^{0})$
        \\
        $--++$ or $-+-+$
        &
        $\mathcal O(g_-^{0})$
        &
        $\mathcal O(g_-^{0})$
        &
        $\mathcal O(g_-^{0})$
        &
        $\mathcal O(g_-^{0})$
        &
        $\mathcal O(g_-^{2})$
        \\
        $---+$
        &
        $\mathcal O(g_-^{2})$
        &
        $\mathcal O(g_-^{2})$
        &
        $\mathcal O(g_-^{2})$
        &
        $\mathcal O(g_-^{2})$
        &
        $\mathcal O(g_-^{4})$
        \\
        $----$
        &
        $\mathcal O(g_-^{4})$
        &
        $\mathcal O(g_-^{4})$
        &
        $\mathcal O(g_-^{4})$
        &
        $\mathcal O(g_-^{4})$
        &
        $\mathcal O(g_-^{6})$
        \\
        \hline
    \end{tabular}
    \caption{Leading behavior of the individual four-point Witten
    diagrams in the limit $g_-\to0$ at fixed $g_+$. The estimates are
    generic and may become softer for special kinematic configurations
    for which the leading helicity contraction vanishes.}
    \label{tab:gminus-scaling}
\end{table}

In Appendix~\ref{sec:Cancellation of divergences in the self-dual limit}, we show that all potential divergences arising in the self-dual limit \eqref{eq:limit_g x} cancel. This is one of the main results of this section. Furthermore, we derive the finite part obtained in the limit after summing all diagrams. We display the final result here:
\begin{equation}
\begin{aligned}
\label{eq:4points}
&\left\langle
J^+(\vec k_1)J^+(\vec k_2)J^+(\vec k_3)J^+(\vec k_4)
\right\rangle
\\
&\quad=
g_+^6\,
\epsilon_i^+(\vec k_1)
\epsilon_j^+(\vec k_2)
\epsilon_k^+(\vec k_3)
\epsilon_l^+(\vec k_4)\,
\epsilon_{ijm}\epsilon_{kln}
\left[
\frac{\Pi^+_{mn}(\vec q)}{q}
-\frac{E+q}{E_LE_R}\Pi^-_{mn}(\vec q)
-\frac{E}{k_{12}k_{34}}L_{mn}(\vec q)
\right]
+\bigl(2\leftrightarrow4\bigr)\,,
\\[0.8em]
&\left\langle
J^-(\vec k_1)J^+(\vec k_2)J^+(\vec k_3)J^+(\vec k_4)
\right\rangle
\\
&\quad=
g_+^6\,k_1\,
\epsilon_i^-(\vec k_1)
\epsilon_j^+(\vec k_2)
\epsilon_k^+(\vec k_3)
\epsilon_l^+(\vec k_4)\,
\epsilon_{ijm}\epsilon_{kln}
\left[
\frac{\Pi^+_{mn}(\vec q)}{E_L}
\left(
\frac{1}{q}+\frac{1}{E}
\right)
-\frac{\Pi^-_{mn}(\vec q)}{EE_R}
-\frac{L_{mn}(\vec q)}{Ek_{34}}
\right]
+\bigl(2\leftrightarrow4\bigr)\,,
\\[0.8em]
&\left\langle
J^-(\vec k_1)J^+(\vec k_2)J^-(\vec k_3)J^+(\vec k_4)
\right\rangle =
g_+^6\,k_1k_3\,
\epsilon_i^-(\vec k_1)
\epsilon_j^+(\vec k_2)
\epsilon_k^-(\vec k_3)
\epsilon_l^+(\vec k_4)\,
\epsilon_{ijm}\epsilon_{kln}\,
\frac{\Pi^+_{mn}(\vec q)}
{qE_LE_R}
+\bigl(2\leftrightarrow4\bigr)\,,
\\[0.8em]
&\left\langle
J^-(\vec k_1)J^+(\vec k_2)J^+(\vec k_3)J^-(\vec k_4)
\right\rangle =
g_+^6\,k_1k_4\,
\epsilon_i^-(\vec k_1)
\epsilon_j^+(\vec k_2)
\epsilon_k^+(\vec k_3)
\epsilon_l^-(\vec k_4)\,
\epsilon_{ijm}\epsilon_{kln}\,
\frac{\Pi^+_{mn}(\vec q)}
{qE_LE_R}\,,
\\[0.8em]
&\left\langle
J^-(\vec k_1)J^-(\vec k_2)J^-(\vec k_3)J^+(\vec k_4)
\right\rangle
=
\left\langle
J^-(\vec k_1)J^-(\vec k_2)J^-(\vec k_3)J^-(\vec k_4)
\right\rangle
=0~.  
\end{aligned}
\end{equation}
Here, we have introduced the notations
\begin{equation} \label{eq:4_pt_notation}
\begin{aligned}
& k_a\equiv|\vec k_a|,
\qquad
E\equiv\sum_{a=1}^4k_a,
\\
&\vec q=\vec k_1+\vec k_2,
\qquad
q=|\vec k_1+\vec k_2|,
\qquad
k_{ab}=k_a+k_b,
\\
&E_L=k_{12}+q,
\qquad
E_R=k_{34}+q~.
\end{aligned}
\end{equation}
The notation $(2\leftrightarrow4)$ means that the complete preceding
$s$-channel expression must be repeated after exchanging the external
legs $2$ and $4$. In particular, in the exchanged term,
\begin{equation}
\begin{aligned}
\vec q&\longrightarrow\vec k_1+\vec k_4,
&
q&\longrightarrow|\vec k_1+\vec k_4|,
\\
E_L&\longrightarrow k_1+k_4+|\vec k_1+\vec k_4|,
&
E_R&\longrightarrow k_2+k_3+|\vec k_1+\vec k_4|~.
\end{aligned}
\end{equation}

\subsection{Correlators of $\mathcal J$}
\label{sec:CorrelatorsofJ}

We now present the corresponding correlators of the full operator \eqref{boundary current}. We derive them in Appendix \ref{sec:Correlators of the non-linear operator} by adding the non-linear terms to \eqref{boundary current lin}. As explained there, these terms contribute to tree-level correlators through appropriate Wick contractions. Again, the terms that are potentially divergent in the self-dual limit are shown to cancel, yielding only finite contributions. Here, we write the final result for the tree-level, color-ordered self-dual correlators up to four points. At two points, we get
\begin{equation} \label{2ptsdfull}
     \left\langle\mathcal J_i^a(\vec k)
     \mathcal J_j^b(-\vec k)\right\rangle
     =\delta^{ab}k
     g_+^2\Pi^+_{ij}(\vec k) \;.
\end{equation}
At three points, we find
\begin{align}
\left\langle\mathcal J^+(\vec k_1)
\mathcal J^+(\vec k_2)\mathcal J^+(\vec k_3)\right\rangle
&=-2g_+^4\epsilon_{ijk}
 \epsilon_i^+(\vec k_1)\epsilon_j^+(\vec k_2)
 \epsilon_k^+(\vec k_3),
\nonumber\\
\left\langle\mathcal J^-(\vec k_1)
\mathcal J^+(\vec k_2)\mathcal J^+(\vec k_3)\right\rangle
&=-g_+^4  \frac{k_2+k_3}{E}
 \epsilon_{ijk}\epsilon_i^-(\vec k_1)
 \epsilon_j^+(\vec k_2)\epsilon_k^+(\vec k_3),
\nonumber\\
\left\langle\mathcal J^-(\vec k_1)
\mathcal J^-(\vec k_2)\mathcal J^+(\vec k_3)\right\rangle
&=0,
\nonumber\\
\left\langle\mathcal J^-(\vec k_1)
\mathcal J^-(\vec k_2)\mathcal J^-(\vec k_3)\right\rangle
&=0.
\label{eq:full-current-three-point-SD}
\end{align}
The four-point correlators are given by
\begin{align}
    &\left\langle
    \mathcal J^+(\vec k_1)\mathcal J^+(\vec k_2)
    \mathcal J^+(\vec k_3)\mathcal J^+(\vec k_4)
    \right\rangle  =g_+^6
    \epsilon_i^+(\vec k_1)\epsilon_j^+(\vec k_2)
    \epsilon_k^+(\vec k_3)\epsilon_l^+(\vec k_4)
    \epsilon_{ijm}\epsilon_{kln}
    \nonumber\\
    &\qquad\qquad\qquad\qquad\qquad\qquad \times
    \left[
    \frac{\Pi^+_{mn}(\vec q)}{q}
    +\frac{E+3q}{E_LE_R}\Pi^-_{mn}(\vec q)
    +\frac{E}{k_{12}k_{34}}L_{mn}(\vec q)
    \right]
    +(2\leftrightarrow4),
    \nonumber \\
    &\left\langle
    \mathcal J^-(\vec k_1)\mathcal J^+(\vec k_2)
    \mathcal J^+(\vec k_3)\mathcal J^+(\vec k_4)
    \right\rangle = g_+^6
    \epsilon_i^-(\vec k_1)\epsilon_j^+(\vec k_2)
    \epsilon_k^+(\vec k_3)\epsilon_l^+(\vec k_4)
    \epsilon_{ijm}\epsilon_{kln}
    \nonumber\\
    &\qquad\qquad\qquad\qquad\qquad\qquad
    \times\!\left[
    \frac{\Pi^+_{mn}(\vec q)}{E_L}
    \!\left(\frac{k_1}{E}+\frac{k_2+q}{q}\right)
    +\frac{E-k_1}{E}
    \!\left(\frac{\Pi^-_{mn}(\vec q)}{E_R}
    +\frac{L_{mn}(\vec q)}{k_{34}}\right)
    \right]\!
    +(2\leftrightarrow4),
    \nonumber \\
    &\left\langle
    \mathcal J^-(\vec k_1)\mathcal J^+(\vec k_2)
    \mathcal J^-(\vec k_3)\mathcal J^+(\vec k_4)
    \right\rangle
    =g_+^6
    \epsilon_i^-(\vec k_1)\epsilon_j^+(\vec k_2)
    \epsilon_k^-(\vec k_3)\epsilon_l^+(\vec k_4)
    \epsilon_{ijm}\epsilon_{kln}\nonumber \\
&\qquad\qquad\qquad\qquad\qquad\qquad\qquad\qquad\qquad\qquad\qquad\qquad
   \times
    \frac{(k_2+q)(k_4+q)}{qE_LE_R}\Pi^+_{mn}(\vec q)
    +(2\leftrightarrow4),
    \nonumber\\
    &\left\langle
    \mathcal J^-(\vec k_1)\mathcal J^+(\vec k_2)
    \mathcal J^+(\vec k_3)\mathcal J^-(\vec k_4)
    \right\rangle
    =g_+^6
    \epsilon_i^-(\vec k_1)\epsilon_j^+(\vec k_2)
    \epsilon_k^+(\vec k_3)\epsilon_l^-(\vec k_4)
    \epsilon_{ijm}\epsilon_{kln}  \frac{(k_2+q)(k_3+q)}{qE_LE_R}\Pi^+_{mn}(\vec q),
    \nonumber\\
    &\left\langle\mathcal J^-\mathcal J^-\mathcal J^-\mathcal J^+\right\rangle
    =
    \left\langle\mathcal J^-\mathcal J^-\mathcal J^-\mathcal J^-\right\rangle=0. \label{4ptsdfull}
\end{align}
Importantly, as explained in Appendix \ref{sec:Correlators of the non-linear operator}, the total energy poles are the same as for the correlators of the linearized operator discussed in the previous section. Additionally, the all-minus and single-plus correlators still vanish.

In summary, we have
shown that, through four points,  the correlators remain finite and simplify drastically in the self-dual limit \eqref{eq:limit_g x}  with a chiral helicity structure. The computation required nontrivial
cancellations among diagrams that diverge individually. In the next
section, we provide further evidence that the limit \eqref{eq:limit_g x}
indeed describes SDYM by rewriting the theory in its
first-order BF formulation. In the latter formulation, the self-dual
limit is manifest and can be taken diagram by diagram, with every
diagram remaining finite.

\section{BF formulation}
\label{sec:BF formulation}

In this section, we study 
the first-order BF formulation of self-dual Yang-Mills theory in AdS$_4$. We find that it requires a boundary condition which relates $B$ to $F$, and breaks the SO(5,1)   conformal group to the SO(4,1) AdS$_4$ isometry group. We will also see that the $BF $ definition of self-dual Yang--Mills theory in AdS$_4$ agrees perturbatively with 
the self-dual limit defined by \eqref{eq:limit_g x}.  This allows us to work directly at the self-dual point, where the computations simplify.  Correlation functions of the resulting theory can be computed perturbatively using standard Witten diagrams.

\subsection{Self-dual limit in the BF formulation}
\label{sec:Self-dual limit in the BF formulation}

The standard approach to defining the self-dual limit of Yang--Mills theory in flat space uses the first-order BF formulation \cite{Chalmers:1996rq}. Here we extend this to AdS$_4$, which entails boundary conditions. For full Yang-Mills, one introduces a dynamical anti-self-dual two-form $B_{\mu \nu}^a$ and considers the theory defined by the action
\begin{equation} \label{eq:BF_action}
	S_{BF} [A, B] = \int_{{\rm AdS}_4} d^4x\,\sqrt{g}\,
\left[
 B^a_{\mu\nu}F^{-,a\,\mu\nu}
-
\frac{g_+^2g_-^2}{g_+^2+g_-^2}
 B^a_{\mu\nu} B^{a\,\mu\nu}
+
\frac{1}{4g_+^2}
F^a_{\mu\nu}\star F^{a\,\mu\nu}
\right]~.
\end{equation} This first-order theory is equivalent to full, non self-dual Yang--Mills theory in the
presence of a $\theta$ term. Starting from
\begin{equation}
\label{path integral BF}
Z_{\mathrm{BF}}
=
\int \mathcal D B\,\mathcal D A\, e^{-S_{BF}[A,B]}~,
\end{equation}
and performing the Gaussian integral over $B$ along the appropriate
contour, whose stationary value is
\begin{equation}
\label{EOMB}
B^a_{\mu\nu}
=
\frac{1}{g^2}F^{-,a}_{\mu\nu}~,
\end{equation}
we recover the standard Yang-Mills path integral \eqref{second order path integral}, up to a
field-independent normalization.

Since $B$ is an auxiliary field fixed algebraically by its
equation of motion, the boundary conditions in the first-order
formulation follow from the second-order boundary conditions
\eqref{eq:N_F}. They are given by
\begin{equation}
\label{eq:N_BF}
n^\mu
\left(
B_{\mu\nu}^a
+
\frac{1}{2g_+^2}\star F_{\mu\nu}^a
\right)
\Big|_{\partial\mathrm{AdS}_4}
=0~.
\end{equation}
They can also be understood directly from the BF path integral as the equations of motion for the boundary value $A_i\big|_{z=0}$, which is path-integrated over. These boundary conditions render the BF action
\eqref{eq:BF_action} stationary on shell.

The self-dual theory is obtained by keeping $B$ and $A$ finite and
taking the limit \eqref{eq:limit_g x}. In this limit, the classical equation of motion \eqref{EOMB} imposes \eqref{eq:SD}. In the path
integral, the limiting theory is simply
\begin{equation}
Z_{\mathrm{BF}}^{\mathrm{sd}}
=
\int \mathcal D B\,\mathcal D A\,
\exp\left\{
-\int_{{\rm AdS}_4} d^4x\,\sqrt{g}\,
\left[
B^a_{\mu\nu}F^{-,a\,\mu\nu}
+
\frac{1}{4g_+^2}
F^a_{\mu\nu}\star F^{a\,\mu\nu}
\right]
\right\}~.
\end{equation}
The field $B$ now acts as a Lagrange multiplier and, when integrated
out with the appropriate contour prescription, produces
$\delta[F^-]$. We therefore recover the (singular) self-dual path integral
\eqref{sd path integral} obtained by taking the self-dual limit in the
second-order formulation. Thus, up to a field-independent
normalization and with matched boundary conditions, the two
formulations, as well as their self-dual limits, formally agree even at the quantum
level.  Below, we verify this
equivalence explicitly at tree level and show that the limit of BF correlators is smooth diagram by diagram, unlike those of the second-order formalism. 

As mentioned earlier, $g_+$ is a genuine coupling constant of the self-dual theory. If we canonically normalize the operators to have $O(1)$ two-point functions, then higher-point correlators are suppressed by powers of $g_+$. $g_+$ is the loop-counting parameter of the self-dual theory. This can be seen by rescaling $B = \frac{1}{g_+^2} \tilde B$, giving the self-dual BF action
\begin{equation} \label{eq:loop_counting}
    S = \frac{1}{g_+^2} \int_{{\rm AdS}_4} d^4x\,\sqrt{g}\,
    \left[
    \tilde B^a_{\mu\nu}F^{-,a\,\mu\nu}
    +
    \frac{1}{4}
    F^a_{\mu\nu}\star F^{a\,\mu\nu}
    \right] \;.
\end{equation}
Although we do not consider non-perturbative effects herein, this expression suggests that they go like powers of $e^{-1/g_+^2}$.

\subsection{Propagators, vertices, and their self-dual limit}

In the following, we work in Poincaré coordinates and fix the radial-gauge condition \eqref{radial gauge}. Since $B_{\mu\nu}$ is an anti-self-dual field, it has three components, which we define by
\begin{equation}
     b_i \equiv B_{zi}~.
\end{equation} Anti-self-duality $B_{\mu\nu} = -\star B_{\mu\nu} = - \frac{1}{2} \epsilon_{\mu\nu\rho\sigma} B^{\rho\sigma}$ then further implies
\be
\qquad B_{ij} = - \epsilon_{ijk} b_k~.
\ee

The bulk-to-bulk propagators are Green's functions for the quadratic operator in the BF action \eqref{eq:BF_action}. In Fourier space at the boundary of AdS, the propagators regular in the interior and compatible with the Neumann boundary conditions \eqref{eq:N_BF} read as 
\begin{align}
         \left\langle A_i^a(\vec k,z)A_j^b(-\vec k,z')\right\rangle &= \frac{g_+^2 g_-^2}{g_+^2+g_-^2} \delta^{ab} \frac{1}{k} \left[ \Pi^+_{ij} \left( e^{-k|z-z'|} + \frac{g_+^2}{g_-^2} e^{-k(z+z')} \right) \right. \nonumber\\
         &\qquad\qquad\qquad\qquad\left. + \Pi^-_{ij} \left( e^{-k|z-z'|} + \frac{g_-^2}{g_+^2} e^{-k(z+z')} \right) + 2 k \,L_{ij} (l(k) - \max(z,z')) \right] \, , \nonumber \\
         \left\langle A_i^a(\vec k,z)b_j^b(-\vec k,z')\right\rangle &=  \frac{1}{2}\delta^{ab} \left[ \Pi^+_{ij} \theta(z-z') e^{-k|z-z'|} \right. \nonumber
         \\
         &\qquad\qquad \left.  -\Pi^-_{ij} \left( \theta(z'-z) e^{-k|z-z'|} + \frac{g_-^2}{g_+^2} e^{-k(z+z')}
        \right) -L_{ij} \theta(z'-z) \right] \, , \nonumber \\
         \left\langle b_i^a(\vec k,z)b_j^b(-\vec k,z')\right\rangle &= \delta^{ab} k   \Pi^-_{ij} \frac{g_+^2+g_-^2}{4g_+^4} e^{-k(z+z')}~.
\label{eq:B2Bafterboundarycond}
\end{align}
We refer to Appendix~\ref{sec:Propagators in BF theory} for details of the derivation of these propagators. One can check their mutual compatibility using the linearized equation of motion \eqref{EOMB}, which, in momentum space, reads
\begin{equation}
     b^a_i (k, z) = \frac{1}{2g^2} (\partial_z \delta_{ik} - i \epsilon_{ijk} k_j) A^a_k (k,z)~, 
\end{equation} together with the Schwinger--Dyson identity. In particular, we have the useful relation
\begin{equation} \label{Schwinger-Dyson}
\left\langle
b_i^a(\vec k,z)b_j^b(-\vec k,z')
\right\rangle
=
\frac{1}{g^4}
\left\langle
F_{zi}^{-,a}(\vec k,z)
F_{zj}^{-,b}(-\vec k,z')
\right\rangle
-
\frac{1}{4g^2}\,
\delta^{ab}\delta_{ij}\,
\delta(z-z')~. 
\end{equation}
The bulk-to-bulk propagators \eqref{eq:B2Bafterboundarycond} are finite in the self-dual limit \eqref{eq:limit_g x} and reduce to
\begin{align}
         \left\langle A_i^a(\vec k,z)A_j^b(-\vec k,z')\right\rangle &= g_+^2\delta^{ab} \frac{1}{k} \Pi_{ij}^+  e^{-k(z+z')} \, , \nonumber  \\
         \left\langle A_i^a(\vec k,z)b_j^b(-\vec k,z')\right\rangle &= \frac{1}{2} \delta^{ab} e^{-k|z-z'|} \left[ \Pi^+_{ij} \theta(z - z') -\Pi^-_{ij}
        \theta(z' - z) \right] - \frac{1}{2}\delta^{ab} L_{ij} \theta(z' - z) \, , \nonumber \\
        \left\langle b_i^a(\vec k,z)b_j^b(-\vec k,z')\right\rangle & = \frac{k}{4 g_+^2} \delta^{ab}   \Pi^-_{ij} e^{-k(z+z')}~. 
        \label{eq:B2Bsd}
\end{align} 
Note that $F^-$ insertions are zero everywhere in the bulk away from coincident
$b$ insertions: $\langle F^- A\rangle = 0$ and $\langle F^-(\vec{x},z) b(\vec{y},z')\rangle \propto \delta^3(\vec{x}-\vec{y}) \delta(z-z')$. Taking $z\rightarrow 0$ in \eqref{eq:B2Bsd} and applying the operator \eqref{boundary current lin} to the boundary insertion of the gauge field, we find the self-dual bulk-to-boundary propagators:
\begin{equation}
\label{eq:B2b}
         \langle J^a_i (\vec k) A^b_j(- \vec k,z')\rangle = \delta^{ab} g_+^2\, e^{-kz'} \Pi^+_{ij} , \qquad
         \langle J^a_i(\vec k) b^b_j(- \vec k,z')\rangle = \frac{k}{2} \delta^{ab} e^{-kz'}
        \Pi^-_{ij}~.
\end{equation}
Notice that, because of \eqref{eq:N_BF}, the boundary value of the gauge field and the anti-self-dual two-form are related by $b_i |_{z=0} = - \frac{1}{2 g_+^2} \star F_{zi} |_{z=0}$. Therefore, there is only one independent boundary current, the same as in the second-order formulation \eqref{boundary current}.

Equation~\eqref{eq:B2b} leads to an interesting conclusion. Although it is
common to interpret $A$ and $B$ as the positive- and negative-helicity
fields, respectively, in the bulk BF theory \eqref{eq:BF_action}, it is not
a priori obvious that this notion coincides, from the boundary perspective,
with the notion of helicity used in the three-dimensional CFT literature
\cite{Caron-Huot:2021kjy}. There, helicity is defined by the eigenvalues $\pm1$ of the operator
$h_{ij}=\frac{i}{k}\epsilon_{ijm}k^m$ introduced in
\eqref{projectors}. Equation~\eqref{eq:B2b}
shows that
$\langle J^a_i(\vec k) A^b_j(-\vec k,z')\rangle$ and
$\langle J^a_i(\vec k) b^b_j(-\vec k,z')\rangle$
belong to its $+1$ and $-1$ eigenspaces, respectively, since
$h_{ik}\Pi^\pm_{kj}=\pm\Pi^\pm_{ij}$. Thus, the bulk and boundary notions
of helicity agree.

The action \eqref{eq:BF_action} leads only to cubic interaction vertices. The BF term contains a cubic interaction involving one $b$ field and
two gauge fields. For the ordering $b_i^a A_j^b A_k^c$, the
corresponding Witten-diagram vertex is
\begin{equation}
    \left(V^{\mathrm{BF}}_3\right)_{ijk}^{abc}
    =
    2 \epsilon_{ijk}f^{abc}
    \int_0^\infty dz~.
\end{equation}
The topological $F\wedge F$ term also contains a cubic interaction.
Since this term reduces to a boundary Chern--Simons functional, the
corresponding boundary vertex is
\begin{equation}
    \left(V^{\mathrm{CS}}_3\right)_{ijk}^{abc}
    = 
    \frac{1}{g_+^2}\epsilon_{ijk}f^{abc}~,
\end{equation}
and is evaluated at $z=0$. 

\begin{figure}[h!]
    \centering
\includegraphics[width=1.0\textwidth]{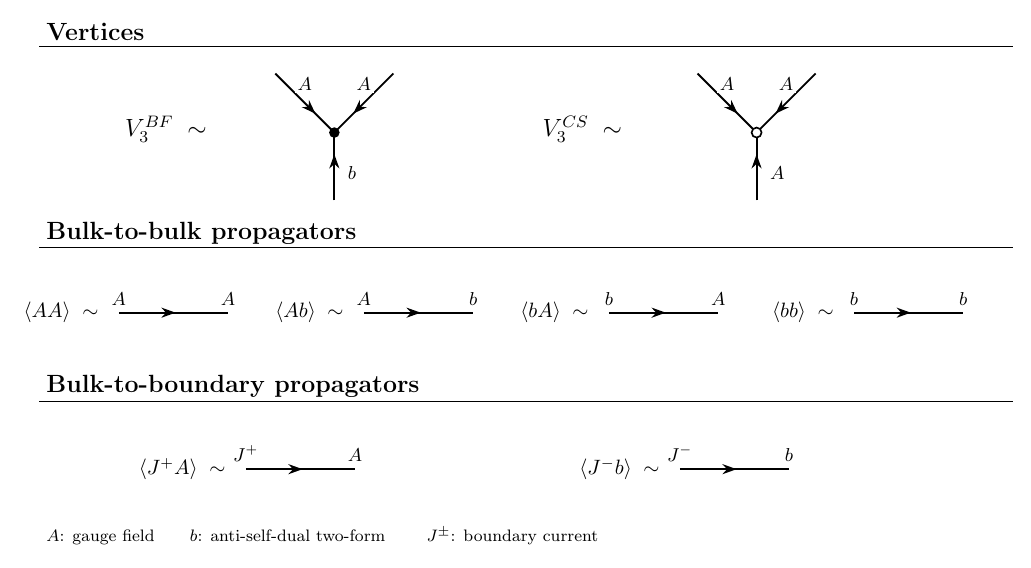}
\caption{Rules for Witten diagrams associated with the self-dual limit of the BF theory \eqref{eq:BF_action}.}
\label{fig:Feynman}
\end{figure}

In contrast to the vertices of the second-order formulation, those of
the first-order BF formulation remain the same in the self-dual limit
\eqref{eq:limit_g x}. Only the propagators vary as we tune $g_-$, but we showed above that they have smooth, finite limits as we approach the self-dual point \eqref{eq:limit_g x}. Consequently, all tree-level Witten diagrams are
individually finite in this limit, and only the finite parts of the
bulk-to-bulk propagators \eqref{eq:B2Bsd} and the bulk-to-boundary propagators
\eqref{eq:B2b} need to be retained when computing the self-dual
correlators. The building blocks of the Witten diagrams associated with the self-dual BF theory are summarized in
Figure~\ref{fig:Feynman}.

\subsection{Boundary correlators}
\label{sec:Self-dual holographic correlators2}

The two-point function is obtained by taking $z' \to 0$ in the first equation of \eqref{eq:B2b} and projecting onto $J^b_j (-\vec k )$,\footnote{One could, in principle, also take $z' \to 0$ in the second equation of \eqref{eq:B2b} and use the boundary conditions \eqref{eq:N_BF} to obtain $J^b_j (-\vec k )$ in the second insertion. However, the result would differ from \eqref{2-point} by a contact term. Here, we fix the boundary operator to be \eqref{boundary current lin} from the outset to avoid this type of contact-term ambiguity.} leading to
\begin{equation} \label{2-point}
     \langle J^a_i (\vec k) J^b_j(-\vec k)\rangle = \delta^{ab} g_+^2\,k\, \Pi_{ij}^+~.
\end{equation}
This reproduces the result \eqref{2ptfunction} obtained by taking the self-dual limit in the second-order formulation.

\begin{figure}[h!]
    \centering
\includegraphics[width=1.0\textwidth]{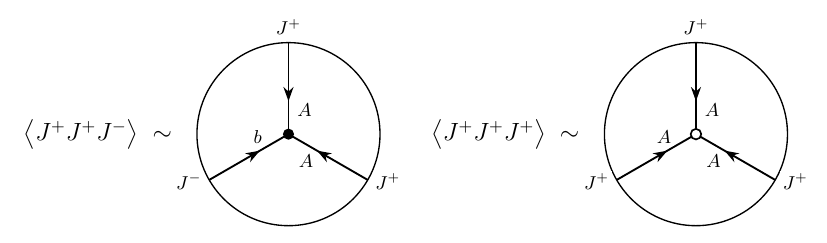}
\caption{Three-point contact diagrams.}
\label{fig:3pt}
\end{figure}

At three points, there are two contact diagrams, displayed in Figure~\ref{fig:3pt}. As stated above, these diagrams are automatically finite in the BF formulation, and their explicit computation reproduces \eqref{3pt correlators}.

\begin{figure}[h!]
    \centering
\includegraphics[width=1.0\textwidth]{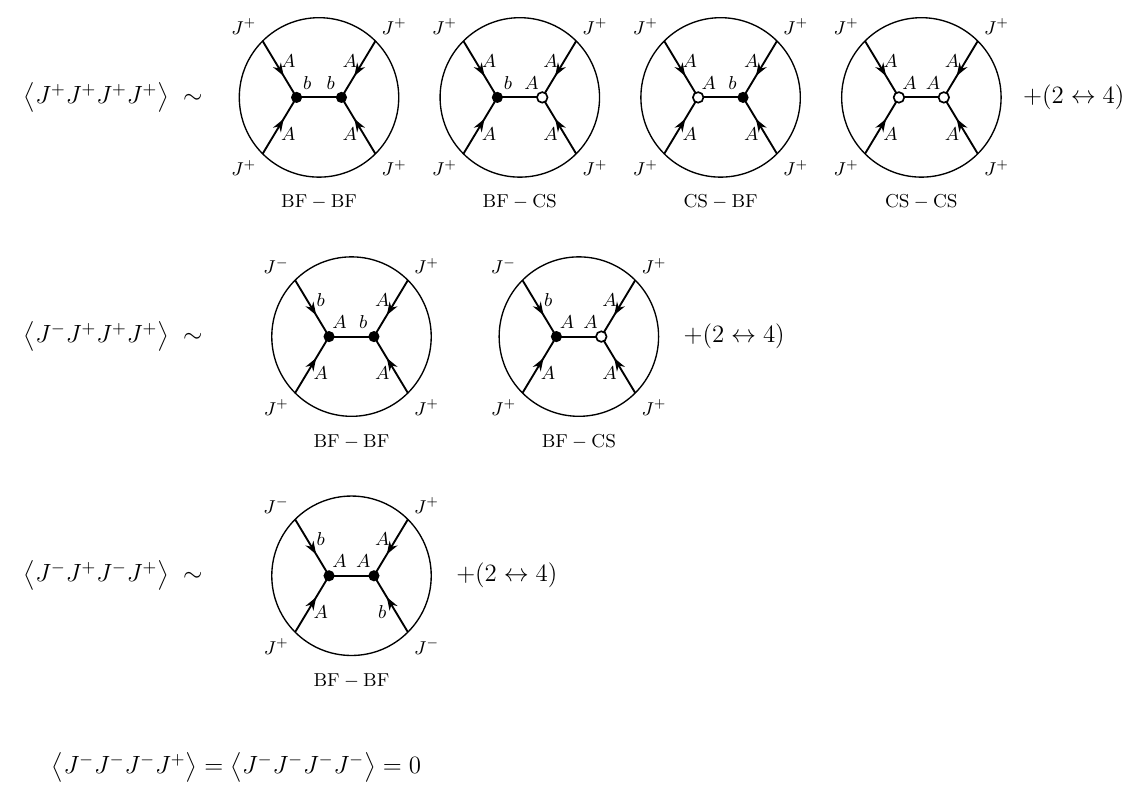}
\caption{Four-point diagrams.}
\label{fig:4pt}
\end{figure}

Finally, since the BF theory contains only cubic vertices, all possible tree-level four-point Witten diagrams are exchange diagrams of the type displayed in Figure~\ref{fig:4pt}. Details of the Witten-diagram computations can be found in Appendix~\ref{sec:Four-point functions in the BF formulation}, and the final results correctly reproduce \eqref{eq:4points}.
For the correlators of the full non-linear operator \eqref{boundary current}, we refer to Appendix \ref{app:nonlinear-current-BF}.

Hence, we have shown that, at tree level, the self-dual limits of
Yang--Mills theory in the first- and second-order formulations agree
and yield a finite, well-defined tree-level realization of self-dual
Yang--Mills theory. This confirms the general path-integral discussion
of Sections~\ref{sec:Self-dual limit of Yang-Mills theory} and \ref{sec:Self-dual limit in the BF formulation}. The two computations use the same definitions
of the boundary operators and the same boundary conditions. To understand this equivalence diagrammatically, one must use the
Schwinger--Dyson identities relating the propagators of the
first-order theory to correlators in the second-order theory; see the
discussion around \eqref{Schwinger-Dyson}. In particular, these
identities express a $BB$ bulk-to-bulk propagator in terms of an
$F^-F^-$ bulk-to-bulk propagator and a local contact term.
Consequently, a $BB$ exchange diagram in the BF formulation
corresponds, in the second-order formulation, to the sum of an
exchange contribution and a contact contribution.

At higher multiplicity $n$, the diagrammatic structure of the vertices and propagators immediately implies that the zero-plus and single-plus correlators vanish for all $n$. The first non-vanishing tree-level $n$-point correlator is the double-plus, $(n-2)$-minus correlator.   We will compute it explicitly at arbitrary multiplicity in the next subsection.

\subsection{All multiplicity self-dual correlator} \label{sec:n pt correlator}
\begin{figure}[h!]
    \centering
\includegraphics[width=0.8\textwidth]{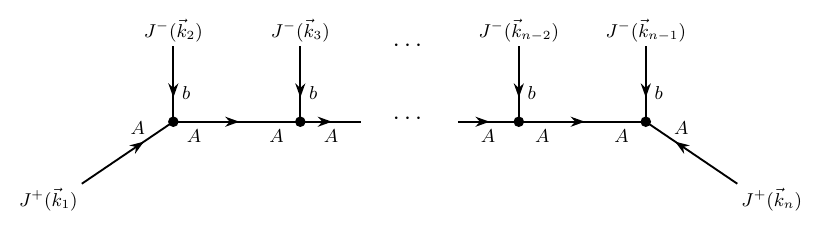}
    \caption{The unique planar ``ladder" diagram contributing to the
    color-ordered 2-plus, $(n-2)$-minus correlator.}
    \label{fig:n-point-BF-ladder}
\end{figure}

We consider the tree-level color-ordered $n$-point correlator of the linearized operator \eqref{boundary current lin},
\begin{equation} \label{eq:}
    \left\langle J^+(\vec{k}_1) J^-(\vec{k}_2) J^-(\vec{k}_3) \dots J^-(\vec{k}_{n-1}) J^+(\vec{k}_n) \right\rangle \;.
\end{equation}
In the self-dual BF theory, it is given by a single planar Witten diagram shown in Figure \ref{fig:n-point-BF-ladder}.

This diagram contains $n-2$ $BAA$ vertices and $n-3$ $\langle AA\rangle$ internal propagators, and can be evaluated explicitly to give
\begin{equation} \label{eq:n pt_correlator}
    \begin{split}
        \bigg\langle J^+(\vec{k}_1) &J^-(\vec{k}_2) J^-(\vec{k}_3) \dots J^-(\vec{k}_{n-1}) J^+(\vec{k}_n) \bigg\rangle \\
        &= g_+^{2n-2} \epsilon_{i_1}^+(\vec{k}_1) \epsilon_{i_2}^-(\vec{k}_2) \dots \epsilon_{i_{n-1}}^-(\vec{k}_{n-1}) \epsilon^+_{i_n} (\vec{k_n})  
        \\
        &\quad 
        \times  \epsilon_{i_1 i_2 j_2} \Pi^+_{j_2 k_2} (\vec{k}_1+\vec{k}_2)  \epsilon_{k_2 i_3 j_3} \Pi^+_{j_3 k_3}(\vec{k}_1 + \vec k_2+\vec k_3) \dots \Pi^+_{j_{n-2} k_{n-2}}(\vec k_1 + \dots + \vec k_{n-2}) \epsilon_{k_{n-2} i_{n-1} i_n} \\
        &\quad \times \left( \prod_{j=2}^{n-2} \frac{1}{\left| \sum_{i=1}^j \vec k_i \right|} \right) \cdot \left(\prod_{j=2}^{n-1} \frac{k_j}{k_j + \left| \sum_{i=1}^{j-1} \vec k_i \right|+ \left| \sum_{i=1}^{j} \vec k_i \right|} \right) \;.
    \end{split}
\end{equation}
The first product in the last line comes from the $n-3$ internal propagators. The denominators in the second product come from the $n-2$ radial integrals in the vertices, and the numerators come from the $\langle J^- b\rangle$ bulk-to-boundary propagators. 

We can also compute the same tree-level correlator using the full boundary operator \eqref{boundary current}. This is computed in Appendix \ref{app:n_pt correlator non lin}, and the final result is given by \eqref{eq:n pt_correlator non_lin}. This is related to the linearized answer above by the simple multiplicative factor
\begin{equation}
    \begin{split}
        \bigg\langle \mathcal J^+(\vec{k}_1) & \mathcal J^-(\vec{k}_2)  \dots \mathcal J^-(\vec{k}_{n-1}) \mathcal J^+(\vec{k}_n) \bigg\rangle \\
        &= \left( \prod_{j=2}^{n-1} \frac{-\left| \sum_{i=1}^{j-1} \vec k_i \right|- \left| \sum_{i=1}^{j} \vec k_i \right|}{k_j} \right) \bigg\langle J^+(\vec{k}_1) J^-(\vec{k}_2)\dots J^-(\vec{k}_{n-1}) J^+(\vec{k}_n) \bigg\rangle \;.
    \end{split}
\end{equation}

Note that this correlator does not have a total energy pole for $n> 3$.\footnote{At $n=3$, this coincides with the single-minus correlator, which has a total energy pole.} As discussed in Section \ref{sec:Single-minus gluon amplitudes}, this is compatible with the vanishing of the corresponding flat space amplitude in SDYM.

\subsection{Comparison with flat space propagators}

In this section, we briefly compare some aspects of BF theory between AdS and flat space. Yang--Mills theory is conformally invariant at the classical level, both
in the second-order and BF formulations. In Poincaré coordinates
\eqref{eq:Poinc_metric}, Euclidean AdS$_4$ is conformal to the upper
half-space, $z>0$, of flat space. Hence, the tree-level propagators of the
BF theory can equivalently be computed using the flat metric
\begin{equation}
    ds^2=dz^2+d\vec{x}^{\,2}\,.
\end{equation}
The difference is that Euclidean propagators on the full flat space are
defined by demanding regularity as $z\to\pm\infty$, whereas the AdS
propagators are only required to be regular as $z\to\infty$.
Additionally, the AdS propagators are required to satisfy the boundary
conditions \eqref{eq:N_BF} at $z=0$.

The BF propagators in flat space in radial gauge can be computed by
inverting the quadratic operator \eqref{eq:B2B} and demanding regularity
as $z\to\pm\infty$. We obtain
\begin{align}
    \left\langle
    A_i^a(z,\vec{k})A_j^b(z',-\vec{k})
    \right\rangle
    &= \delta^{ab} \frac{g^2}{2k} [\Pi_{ij} \, e^{-k|z-z'|} -L_{ij} \;k |z-z'|] \,,
    \nonumber\\
    \left\langle
    A_i^a(z,\vec{k})b_j^b(z',-\vec{k})
    \right\rangle
    &=
    \frac{1}{2}\delta^{ab}e^{-k|z-z'|}
    \left[
    \Pi^+_{ij}\theta(z-z')
    -\Pi^-_{ij}\theta(z'-z)
    \right]
    -\frac{1}{2}\delta^{ab}L_{ij}
    \left[
    \theta(z'-z)-\frac{1}{2}
    \right]\,,
    \nonumber\\
    \left\langle
    b_i^a(z,\vec{k})b_j^b(z',-\vec{k})
    \right\rangle
    &=0\,. \label{flat space BF}
\end{align}
Notably, these propagators differ from the AdS propagators \eqref{eq:B2Bafterboundarycond}. The AdS propagators contain an additional dyonic image charge at $-z'$, which imposes the boundary condition \eqref{eq:N_BF} by sourcing the extra $e^{-k|z+z'|}$ terms in the propagator. Upon restricting to AdS $z>0$, this becomes the homogenous solution $e^{-k(z+z')}$, which is required to be regular only for $z \to +\infty$.

The restriction of the correlators \eqref{flat space BF} to the $z=0$ surface in flat space gives the two-point function
\begin{equation}
    \left\langle J_i^a(\vec{k}) J_j^b(-\vec{k}) \right\rangle \equiv \frac{1}{4} \epsilon _{ilm} \epsilon_{jno} \left\langle F_{lm}^a(0,\vec{k}) F_{no}^b(0,-\vec{k}) \right\rangle = \delta^{ab} \frac{g^2}{2} k \, \Pi_{ij}(\vec{k}) \;,
\end{equation} 
which should be contrasted with \eqref{eq:JJ-full-YM}.

In the self-dual limit $g_- \to 0$, \eqref{flat space BF} simplifies to
\begin{align}
    \left\langle
    A_i^a(z,\vec{k})A_j^b(z',-\vec{k})
    \right\rangle
    &= 0 \,,
    \nonumber\\
    \left\langle
    A_i^a(z,\vec{k})b_j^b(z',-\vec{k})
    \right\rangle
    &=
    \frac{1}{2}\delta^{ab}e^{-k|z-z'|}
    \left[
    \Pi^+_{ij}\theta(z-z')
    -\Pi^-_{ij}\theta(z'-z)
    \right]
    -\frac{1}{2}\delta^{ab}L_{ij}
    \left[
    \theta(z'-z)-\frac{1}{2}
    \right]\,,
    \nonumber\\
    \left\langle
    b_i^a(z,\vec{k})b_j^b(z',-\vec{k})
    \right\rangle
    &=0\, ,
\end{align}
which also differ from the AdS self-dual propagators \eqref{eq:B2Bsd} by the same image charges. In particular, the Chern-Simons term is irrelevant in flat-space perturbation theory, so the flat space propagators are independent of $g_+$ at the self-dual point.

Hence, the Euclidean theory on AdS is not obtained by simply
restricting the full flat-space theory to $z>0$.

\section{Single-minus gluon amplitudes from AdS$_4$}
\label{sec:Single-minus gluon amplitudes}

Flat-space amplitudes can be extracted from boundary  correlators in AdS$_4$ using a flat-space limit and an appropriate analytic continuation \cite{Polchinski:1999ry,Giddings:1999jq}. This limit can be implemented in position space \cite{Gary:2009ae,Okuda:2010ym,Maldacena:2015iua,Komatsu:2020sag,Alday:2024yyj}, Mellin space \cite{Penedones:2010ue,Paulos:2016fap,Fitzpatrick:2011hu}, or directly in momentum space \cite{Raju:2012zr,Raju:2012zs,Hijano:2019qmi,Hijano:2020szl,Farrow:2018yni,Gadde:2022ghy,Marotta:2024sce,Marotta:2025qjh}. Here, we focus on the latter procedure.

In this section, we extract the flat-space information contained in the
three- and four-point current correlators of the self-dual theory. The
relevant kinematic limit in momentum space is the zero total-energy limit \cite{Maldacena:2011nz,Raju:2012zr,Raju:2012zs}. 
We associate with
every boundary momentum $\vec k_a$ a complex null four-vector
\begin{equation} \label{eq:k_4_vector}
    k_a^\mu=(ik_a,\vec k_a),
    \qquad
    \delta_{\mu \nu}k_a^\mu k_a^\nu=0~,
\end{equation}
where $k_a=|\vec k_a|$.  The condition
$E=0$ is then  the missing radial momentum-conservation
condition, since
\begin{equation}
    \sum_a k_a^\mu
    =
    \left(
        iE,\sum_a\vec k_a
    \right)~.
\end{equation}
Together with boundary momentum conservation, $E=0$ then restores
four-dimensional momentum conservation. The residue of the
total-energy pole of the AdS correlator gives the corresponding
flat-space amplitude, up to the overall normalization of the external
bulk-to-boundary wavefunctions
\cite{Maldacena:2011nz,Raju:2012zr,Raju:2012zs}. This relation is a
well-established feature of cosmological correlators and a key
ingredient of the cosmological bootstrap
\cite{Arkani-Hamed:2018kmz}.

The half-collinear kinematics relevant for the single-minus amplitudes\footnote{Our conventions for $\epsilon^\pm$ are opposite to the flat-space helicities, but this is offset by the fact that we compute correlators in an all-incoming convention. Hence, what we call single-minus in our conventions exactly matches the flat-space single-minus amplitude in an all-outgoing convention.}
can be accessed directly in this complexified momentum space.
Alternatively, they admit a real realization in Klein signature \cite{Guevara:2026qzd}.

\subsection{Three-point amplitude}

From \eqref{3pt correlators}, we see immediately that only $\langle J^-(\vec k_1)J^+(\vec k_2)J^+(\vec k_3)\rangle$ has a total-energy singularity. We can rewrite it as
\begin{equation}
\begin{split}
 \langle J^-(\vec k_1)J^+(\vec k_2)J^+(\vec k_3)\rangle &=g_+^4\frac{1}{2E}(k_1-k_2-k_3+E)\epsilon_{ijk} \epsilon^-_i(\vec k_1)\epsilon^+_j(\vec k_2)\epsilon^+_k(\vec k_3)\\
 &=g_+^4\frac{1}{2E}(k_1-k_2-k_3)\epsilon_{ijk} \epsilon^-_i(\vec k_1)\epsilon^+_j(\vec k_2)\epsilon^+_k(\vec k_3) +\mathcal{O}(E^0)~.
\end{split}
\end{equation}
The first term is precisely the well-known Yang--Mills three-point function whose total-energy pole yields the correct $-++$ flat-space amplitude.\footnote{The factor of $g_+^4$ is due to the normalization of the wavefunction. If we normalize our wavefunctions so that the scattering states are $e^{i k\cdot x} \epsilon^\pm$, we obtain a $g_+$-independent answer, as appropriate for the flat-space limit. The same remarks apply at four points.} This can be seen most explicitly in the spinor-helicity variables of Appendix~\ref{App:spinorHelicity} \cite{Maldacena:2011nz}.

In full Yang--Mills theory on AdS$_4$, both $-++$ and $+--$ amplitudes are obtained as total-energy poles \cite{Maldacena:2011nz}, as expected from the three-point amplitudes of full Yang--Mills theory in flat space. Here, we obtain only $-++$, which is the only nonzero three-point amplitude of SDYM theory in flat space.

\subsection{Four-point amplitude} 

Again, from \eqref{eq:4points}, we see directly that the only helicity configuration with a potential total-energy pole is the single-minus configuration $\langle J^-(\vec k_1)J^+(\vec k_2)J^+(\vec k_3)J^+(\vec k_4)\rangle$. In this subsection, we show that its total-energy pole vanishes for generic boundary momenta. However, when all external three-momenta are half-collinear, in the sense that $\langle ij\rangle=0$ for all $i,j \in \{1,2,3,4\}$, the total-energy pole becomes singular. This singular result can be regularized by a suitable $i\epsilon$ prescription, which shows that the total-energy pole is precisely the single-minus amplitude supported in the half-collinear regime \cite{Guevara:2026qzd}.

\paragraph{Generic external momenta}

First, let us compute the total-energy pole for generic external momenta. It is easy to see that
\begin{equation}
    \begin{split}
&\langle J^-(\vec k_1)J^+(\vec k_2)J^+(\vec k_3)J^+(\vec k_4)\rangle\\
&=\frac{g_+^6}{E}  \, k_1 \epsilon^-_{i}(\vec k_1) \epsilon^+_{j}(\vec k_2)\epsilon^+_{k}(\vec k_3)\epsilon^+_{l}(\vec k_4) \epsilon_{ijm}\epsilon_{kln}K_{mn}(k_i)+(2\leftrightarrow 4) +\mathcal{O}(E^0)~,
\end{split}
\end{equation}
where we define
\be
K_{mn}(k_i)=\frac{1}{q^2-k_{12}^2}\bigg(i \epsilon_{mno}q_o-k_{12}\delta_{mn}+\frac{q_mq_n}{k_{12}}\bigg)~,
\ee
and recall $k_{12}=k_1+k_2$. Next, using $\vec{q}=\vec{k}_1+\vec{k}_2$, we contract this expression with the relevant polarization and simplify it to
\bea
\epsilon_{i}^-(\vec k_1)
\epsilon_{j}^+(\vec k_2) \epsilon_{ijm}K_{mn}(k_i)= \frac{2i \epsilon^+(\vec k_2)\cdot \vec k_1} {q^2-k_{12}^2}\epsilon_n^-(\vec k_1)~.
\eea
The full residue of the $s$-channel total-energy pole is then proportional to
\be
\label{eq:ijkBracket}
\frac{2i \epsilon^+(\vec k_2)\cdot \vec k_1} {k_{12}^\mu k_{12\mu}}[1^-3^+4^+]~,
\ee
where, $[1^-3^+4^+]=\epsilon_{ijk}\,\epsilon_i^-(\vec k_1)\,\epsilon_j^+(\vec k_3)\,\epsilon_k^+(\vec k_4)$ as defined in \eqref{eq:three-bracket}, and we get
\be
k_{12}^\mu k_{12\mu}=q^2-k_{12}^2~
\ee
from \eqref{eq:4_pt_notation} and \eqref{eq:k_4_vector}. Using $2(\epsilon_1\cdot \vec k_2) (\epsilon_2\cdot \vec k_1)=k_{12}^\mu k_{12\mu}\epsilon_1\cdot\epsilon_2$, this gives
\be
\frac{i \epsilon^-(\vec k_1)\cdot \epsilon^+(\vec k_2)} { \epsilon^-(\vec k_1)\cdot \vec k_2}[1^-3^+4^+]~,
\ee
so that the total-energy pole is
\be
\label{eq:TotE1}
i \bigg(\frac{ \epsilon^-(\vec k_1)\cdot \epsilon^+(\vec k_2)} { \epsilon^-(\vec k_1)\cdot \vec k_2}[1^-3^+4^+] +\frac{ \epsilon^-(\vec k_1)\cdot \epsilon^+(\vec k_4)} { \epsilon^-(\vec k_1)\cdot \vec k_4}[1^-3^+2^+]\bigg)~.
\ee
In Appendix~\ref{app:flat_limit_general}, we derive the following relation:
\be
\label{eq:magicCancel}
0=\frac{\epsilon^-(\vec k_1)\cdot \vec k_4 k_{12}^\mu k_{12\mu}-\epsilon^-(\vec k_1)\cdot \vec k_2 k_{14}^\mu k_{14\mu}}{\epsilon^-(\vec k_1)\cdot \vec k_2+\epsilon^-(\vec k_1)\cdot \vec k_4}\bigg(\frac{i \epsilon^-(\vec k_1)\cdot \epsilon^+(\vec k_2)} { \epsilon^-(\vec k_1)\cdot \vec k_2}[1^-3^+4^+] +\frac{i \epsilon^-(\vec k_1)\cdot \epsilon^+(\vec k_4)} { \epsilon^-(\vec k_1)\cdot \vec k_4}[1^-3^+2^+]\bigg)~,
\ee
which implies that the total-energy pole vanishes for generic kinematics. Note that \eqref{eq:magicCancel} does not imply that the total-energy pole vanishes if the prefactor of the brackets vanishes, as it does in the half-collinear regime. Let us consider this case separately.

\paragraph{Half-collinear regime}

In the previous paragraph, we found that the amplitude obtained from the total-energy pole is proportional to
\be
k_1\bigg(\frac{2i \epsilon^+(\vec k_2)\cdot \vec k_1} {k_{12}^\mu k_{12\mu}}[1^-3^+4^+]+(2\leftrightarrow 4)\bigg) \delta^3\big(\sum \vec{k}_i\big)\delta(E)~,
\ee
where the first term arises from the $s$-channel and the second from the $t$-channel. We argued that this vanishes for generic kinematics, but we now show that it reproduces the single-minus amplitude of \cite{Guevara:2026qzd} in the half-collinear regime, where $\langle ij\rangle=0$ with $[ij]\neq 0$ fixed.\footnote{We analytically continue to complexified momenta, for which this regime is accessible. The real momenta in this regime are obtained by restricting to Klein signature \cite{Guevara:2026qzd}.} To this end, we regularize the residue using the standard $i\epsilon$ prescription:
\be
\label{eq:iepsEPole}
k_1\bigg(\frac{2i \epsilon^+(\vec k_2)\cdot \vec k_1} {k_{12}^\mu k_{12\mu}+i \epsilon}[1^-3^+4^+]+(2\leftrightarrow 4)\bigg) \delta^3\big(\sum \vec{k}_i\big)\delta(E)~.
\ee
In Appendix~\ref{app:flat_limit_collinear}, we show that \eqref{eq:iepsEPole} is proportional to 
\be
\label{eq:finalamplitude}
-\frac{1}{2}(\text{sg}_{12}\text{sg}_{34} +\text{sg}_{23}\text{sg}_{41})\frac{\langle r1\rangle^5}{\langle r2\rangle\langle r3\rangle\langle r4\rangle} \delta(\langle 12\rangle) \delta(\langle 13\rangle) \delta(\langle 14\rangle) \delta^2\bigg(\sum \langle ri\rangle \tilde{\kappa}_i \bigg)~ \;,
\ee
where the spinor notations are defined in Appendix \ref{App:spinorHelicity}, and $\text{sg}_{ij}$ is defined in \eqref{eq:sgn}. Equation~\eqref{eq:finalamplitude} is precisely the single-minus gluon amplitude $\mathcal{A}_4(1^-,2^+,3^+,4^+)$ found in \cite{Guevara:2026qzd}.

\section*{Acknowledgements}

We thank Ofer Aharony, Roland Bittleston, Daniel Jafferis, Trivko Kukolj, Arthur Lipstein, Lionel Mason, Jeffrey Opreij, Guilherme Pimentel, Atul Sharma, David Skinner, and Evgeny Skvortsov for useful discussions. SH is supported by the Gordon and Betty Moore Foundation and the John Templeton Foundation via the Black
Hole Initiative. RR is supported by the European Union’s Horizon Europe research and innovation programme under the Marie Sk{\polishl}odowska--Curie grant agreement No. 101104845 (UniFlatHolo), hosted at Harvard University and École Polytechnique. ASh and ASt are supported by the Gordon and Betty Moore Foundation and the John Templeton Foundation via the Black
Hole Initiative, DOE grant DE-SC/0007870, and the Simons Collaboration on
Celestial Holography. OpenAI LLMs were used at several stages of this project. All results were author-verified.

\appendix

\section{Four-point functions and their self-dual limit}
\label{sec:Cancellation of divergences in the self-dual limit}

In the second-order theory, some diagrams appear to diverge in the self-dual limit \eqref{eq:limit_g x}. We show that these divergent terms cancel when the diagrams are summed. These divergences occur in the $\langle J^+ J^+ J^+ J^+ \rangle$ and $\langle J^- J^+ J^+ J^+ \rangle$ sectors. We also compute explicitly the finite parts of the diagrams in the self-dual limit. We show only the $s$-channel contributions in this section; their divergent terms cancel independently. The $t$-channel contributions are obtained by exchanging external legs 2 and 4, and their divergent terms cancel similarly.

To compute the diagrams, it is useful to decompose the color-ordered vertices \eqref{cubic}, \eqref{quartic} and \eqref{cubic CS} into the transverse helicity basis as follows. First, we define the symbol
\begin{equation} \label{eq:three-bracket}
    [ 1^{\sigma_1} 2^{\sigma_2} 3^{\sigma_3} ] = \epsilon_{ijk} \epsilon_i^{\sigma_1}(\vec{k}_1) \epsilon_j^{\sigma_2}(\vec{k}_2) \epsilon_k^{\sigma_3}(\vec{k_3})
\end{equation}
for helicities $\sigma_1, \sigma_2, \sigma_3 = \pm$. Then, the three-point vertices in helicity basis are
\begin{equation}
    \mathcal V^{\mathrm{YM}}_{\sigma_1 \sigma_2 \sigma_3} =  \frac{1}{2}
    \left(
    \frac{1}{g_+^2}+\frac{1}{g_-^2}
    \right) [ 1^{\sigma_1} 2^{\sigma_2} 3^{\sigma_3} ]  \left( \sigma_1 k_1 + \sigma_2 k_2 + \sigma_3 k_3 \right) \int_0^\infty dz \;,
\end{equation}
and 
\begin{equation}
    \mathcal V^{\mathrm{CS}}_{\sigma_1 \sigma_2 \sigma_3}
    =
    \frac{1}{2}
    \left(
    \frac{1}{g_+^2}-\frac{1}{g_-^2}
    \right)
    [ 1^{\sigma_1} 2^{\sigma_2} 3^{\sigma_3} ]  \;.
\end{equation}
For the color-ordered 4-point vertex, we define 
\begin{equation}
    [ 1^{\sigma_1} 2^{\sigma_2} 3^{\sigma_3} 4^{\sigma_4}] = (\epsilon_{ijm} \epsilon_{klm} + \epsilon_{lim} \epsilon_{mjk}) \epsilon_i^{\sigma_1}(\vec{k}_1) \epsilon_j^{\sigma_2}(\vec{k}_2) \epsilon_k^{\sigma_3}(\vec{k_3}) \epsilon_l^{\sigma_4}(\vec{k}_4) \;.
\end{equation}
Then, the color-ordered 4-point vertex in helicity basis is given by
\begin{equation}
    \mathcal V^{\mathrm{YM}}_{\sigma_1 \sigma_2 \sigma_3 \sigma_4} = -\frac{1}{2}
    \left(
    \frac{1}{g_+^2}+\frac{1}{g_-^2}
    \right)[ 1^{\sigma_1} 2^{\sigma_2} 3^{\sigma_3} 4^{\sigma_4}] \int_0^\infty dz \;. 
\end{equation}

The radial integrals that appear repeatedly are
\begin{equation}
    \int_0^\infty dz\,e^{-E_Lz}
    =
    \frac{1}{E_L},
    \qquad
    \int_0^\infty dz'\,e^{-E_Rz'}
    =
    \frac{1}{E_R}~,
\end{equation}
as well as the ordered integrals
\begin{align}
\int_{z>z'}dz\,dz'\,
e^{-k_{12}z-k_{34}z'-q(z-z')}
&=
\frac{1}{EE_L},
\nonumber \\
\int_{z<z'}dz\,dz'\,
e^{-k_{12}z-k_{34}z'-q(z'-z)}
&=
\frac{1}{EE_R},
\nonumber \\
\int_{z<z'}dz\,dz'\,
e^{-k_{12}z-k_{34}z'}
&=
\frac{1}{Ek_{34}}~.
\label{eq:BF-ordered-integrals}
\end{align}

\subsection{$\langle J^+ J^+ J^+ J^+  \rangle$}

All of the diagrams contributing to the $\langle J^+J^+J^+J^+\rangle$ correlator diverge as $g_- \to 0$. The diagrams are given by 
\begin{align}
\left.\langle J^+J^+J^+J^+\rangle_s\right|_{\mathrm{YM-YM}} =& \epsilon_i^+(\vec k_1)\epsilon_j^+(\vec k_2)
\epsilon_k^+(\vec k_3)\epsilon_l^+(\vec k_4)
\epsilon_{ijm}\epsilon_{kln} \cdot g_+^8 \frac{g_+^2+g_-^2}{4 g_+^2 g_-^2} \frac{1}{q}
\nonumber\\
&\times
\Bigg\{ \left(\frac{E_L+E_R}{E} + \frac{g_+^2}{g_-^2} \right) \Pi^+_{mn}(\vec{q}) 
\nonumber \\
&\quad 
+ \left(\frac{E_L+E_R}{E} + \frac{g_-^2}{g_+^2} \right) \frac{(k_{12}-q)(k_{34}-q)}{E_L E_R} \Pi^-_{mn}(\vec{q}) 
\nonumber \\
&\quad
+ 2q\left( l(q) - \frac{k_{12}^2+k_{12}k_{34}+k_{34}^2}{E k_{12} k_{34}}\right)L_{mn}(\vec{q})\Bigg\}~.
\end{align}

\begin{align}
\left.\langle J^+J^+J^+J^+\rangle_s
\right|_{\mathrm{YM-CS}+\mathrm{CS-YM}}
=&
\epsilon_i^+(\vec k_1)\epsilon_j^+(\vec k_2)
\epsilon_k^+(\vec k_3)\epsilon_l^+(\vec k_4)
\epsilon_{ijm}\epsilon_{kln} \cdot g_+^8 \frac{g_-^2-g_+^2}{2g_+^2g_-^2} \frac{1}{2q}
\nonumber \\
&\times
\Bigg\{
2\left(1+\frac{g_+^2}{g_-^2} \right)
\Pi^+_{mn}(\vec q)
\nonumber\\
&\quad+
\left(1+ \frac{g_-^2}{g_+^2} \right)
\left[
\frac{k_{12}-q}{E_L}
+\frac{k_{34}-q}{E_R}
\right]
\Pi^-_{mn}(\vec q)
\nonumber\\
&\qquad+
2 q
\left[
\frac{k_{12}l(q)-1}{k_{12}}
+\frac{k_{34}l(q)-1}{k_{34}}
\right]
L_{mn}(\vec q)
\Bigg\}~.
\end{align}

\begin{align}
\left.\langle J^+J^+J^+J^+\rangle_s\right|_{\mathrm{CS-CS}} =&
\epsilon_i^+(\vec k_1)\epsilon_j^+(\vec k_2)
\epsilon_k^+(\vec k_3)\epsilon_l^+(\vec k_4)
\epsilon_{ijm}\epsilon_{kln} \cdot g_+^8 \frac{(g_-^2-g_+^2)^2}{4 g_+^2 g_-^2} \frac{1}{q}
\nonumber \\
&\times
\Bigg[
\frac{1}{g_-^2}
\Pi^+_{mn}(\vec q) +
\frac{1}{g_+^2}
\Pi^-_{mn}(\vec q) + \frac{2 q l(q)}{g_+^2+g_-^2} L_{mn}(\vec{q}) \Bigg]~.
\end{align}

\begin{align}
\left.\langle J^+J^+J^+J^+\rangle_s\right|_{\mathrm{contact}}
={}&
\epsilon_i^+(\vec k_1)\epsilon_j^+(\vec k_2)
\epsilon_k^+(\vec k_3)\epsilon_l^+(\vec k_4)
\epsilon_{ijm}\epsilon_{kln} \cdot (-g_+^8) \frac{g_+^2+g_-^2}
{2g_+^2g_-^2} \frac{1}{E}
\nonumber\\
&\times
\left[
\Pi^+_{mn}(\vec q)
+\Pi^-_{mn}(\vec q)
+L_{mn}(\vec q)
\right]~.
\end{align}

Adding all diagrams, we obtain the correlator at arbitrary $g_+$ and $g_-$:
\begin{equation}
\begin{aligned}
\langle J^+ J^+ J^+ J^+ \rangle_s &= \epsilon_i^+(\vec k_1)\epsilon_j^+(\vec k_2)
\epsilon_k^+(\vec k_3)\epsilon_l^+(\vec k_4)
\epsilon_{ijm}\epsilon_{kln} \cdot g_+^6 \\
&\quad \left\{ \frac{1}{q} \Pi^+_{mn}(\vec{q}) + \frac{-q (E+q)+ \frac{g_-^2}{g_+^2} k_{12} k_{34} }{q E_L E_R} \Pi^-_{mn} (\vec{q}) - \left(\frac{1}{k_{12}} + \frac{1}{k_{34}}- \frac{2g_-^2}{g_+^2+g_-^2}l(q) \right) L_{mn}(\vec{q})\right\}~.
\end{aligned}
\end{equation}
This is finite in the $g_- \to 0$ limit and, after adding the $t$-channel, smoothly approaches \eqref{eq:4points}. Note that all dependence on the residual gauge function $l(q)$ vanishes in this limit.

\subsection{$\langle J^-J^+ J^+ J^+ \rangle$}

Only the $\Pi^+$ parts of the exchange diagrams remain separately divergent. The exact relevant pieces are the following:
\begin{align}
\left.\langle J^-J^+J^+J^+\rangle_s
\right|_{\mathrm{YM-YM},\,\Pi^+}
={}&
-g_+^4\frac{g_+^2+g_-^2}{4q}
(k_2-k_1+q)E_R
\nonumber\\
&\times
\left[
\frac1E\left(\frac1{E_L}+\frac1{E_R}\right)
+\frac{g_+^2}{g_-^2E_LE_R}
\right]
\nonumber\\
&\times
\epsilon_i^-\epsilon_j^+\epsilon_k^+\epsilon_l^+
\epsilon_{ijm}\epsilon_{kln}\Pi^+_{mn}~,
\end{align}
\begin{align}
\left.\langle J^-J^+J^+J^+\rangle_s
\right|_{\mathrm{YM-CS},\,\Pi^+}
={}&g_+^4
\frac{g_+^4-g_-^4}{4g_-^2q}
\frac{k_2-k_1+q}{E_L}
\epsilon_i^-\epsilon_j^+\epsilon_k^+\epsilon_l^+
\epsilon_{ijm}\epsilon_{kln}\Pi^+_{mn}~,
\end{align}
\begin{align}
\left.\langle J^-J^+J^+J^+\rangle_s
\right|_{\mathrm{CS-YM},\,\Pi^+}
={}&
g_+^4\frac{g_+^4-g_-^4}{4g_-^2q}
\epsilon_i^-\epsilon_j^+\epsilon_k^+\epsilon_l^+
\epsilon_{ijm}\epsilon_{kln}\Pi^+_{mn}~,
\end{align}
\begin{align}
\left.\langle J^-J^+J^+J^+\rangle_s
\right|_{\mathrm{CS-CS},\,\Pi^+}
={}&
-g_+^4\frac{(g_-^2-g_+^2)^2}{4g_-^2q}
\epsilon_i^-\epsilon_j^+\epsilon_k^+\epsilon_l^+
\epsilon_{ijm}\epsilon_{kln}\Pi^+_{mn}~.
\end{align}
Upon adding all the contributions above, the divergent parts cancel identically.

The remaining finite piece of the bulk Yang--Mills exchange
diagram gives
\begin{align}
\left.\langle J^-J^+J^+J^+\rangle_s
\right|_{\mathrm{YM-YM}}^{\mathrm{fin}}
={}&
-g_+^6\frac{(E+q)(k_2-k_1+q)}
{2qEE_L}
\Pi^+_{mn}(\vec q)
\nonumber\\
&-g_+^6
\frac{(q-k_{34})(q+k_1-k_2)(E+2q)}
{4qEE_LE_R}
\Pi^-_{mn}(\vec q)
\nonumber\\
&+g_+^6
\frac{(k_1-k_2)
\left[
k_{12}k_{34}E\,l(q)
-\left(k_{12}^2+k_{12}k_{34}+k_{34}^2\right)
\right]}
{2k_{12}^2k_{34}E}
L_{mn}(\vec q)~,
\label{eq:finite-mppp-YMYM}
\end{align}
where we suppressed the overall common polarization structure
\begin{equation}
    \epsilon_i^-(\vec k_1)\epsilon_j^+(\vec k_2)
    \epsilon_k^+(\vec k_3)\epsilon_l^+(\vec k_4)
    \epsilon_{ijm}\epsilon_{kln}~.
\end{equation}
The finite part of the sum of the two mixed exchange diagrams is
\begin{align}
\left.\langle J^-J^+J^+J^+\rangle_s
\right|_{\mathrm{mixed}}^{\mathrm{fin}}
={}&g_+^6
\frac{k_{34}k_2-q(k_1+q)}
{2qE_LE_R}
\Pi^-_{mn}(\vec q)
\nonumber\\
&+g_+^6
\frac{
2k_{12}k_{34}k_2\,l(q)
-k_{12}^2-k_{34}(k_2-k_1)
}{
2k_{12}^2k_{34}
}
L_{mn}(\vec q)~.
\label{eq:finite-mppp-mixed}
\end{align}
There is no finite nonzero $\Pi^+$ contribution from the mixed diagrams. The
Chern--Simons exchange diagram contributes
\begin{align}
\left.\langle J^-J^+J^+J^+\rangle_s
\right|_{\mathrm{CS-CS}}^{\mathrm{fin}}
=
\frac{g_+^6}{2}
\left[
\frac{\Pi^+_{mn}(\vec q)}{q}
-\frac{\Pi^-_{mn}(\vec q)}{2q}
-l(q)L_{mn}(\vec q)
\right]~,
\label{eq:finite-mppp-CSCS}
\end{align}
and the quartic contact diagram gives
\begin{align}
\left.\langle J^-J^+J^+J^+\rangle_s
\right|_{\mathrm{contact}}^{\mathrm{fin}}
=
\frac{g_+^6}{2E}
\left[
\Pi^+_{mn}(\vec q)
+\Pi^-_{mn}(\vec q)
+L_{mn}(\vec q)
\right]~.
\label{eq:finite-mppp-contact}
\end{align}

Their sum simplifies to

\begin{align}
&\left.
\left(
\langle J^-J^+J^+J^+\rangle_{\mathrm{YM-YM}}
+\langle J^-J^+J^+J^+\rangle_{\mathrm{mixed}}
+\langle J^-J^+J^+J^+\rangle_{\mathrm{CS-CS}}
+\langle J^-J^+J^+J^+\rangle_{\mathrm{contact}}
\right)_s
\right|_{\mathrm{fin}}
\nonumber\\
&\qquad=
g_+^6 k_1
\left[
\frac{\Pi^+_{mn}(\vec q)}{E_L}
\left(
\frac1q+\frac1E
\right)
-\frac{\Pi^-_{mn}(\vec q)}{EE_R}
-\frac{L_{mn}(\vec q)}{Ek_{34}}
\right]~.
\label{eq:finite-mppp-sum}
\end{align}
Restoring the polarizations and adding the $t$-channel gives the second result in \eqref{eq:4points}.

\subsection{$\langle J^- J^+ J^- J^+ \rangle$}

For two or more external negative helicities, the factors from the
external propagators,
\begin{equation}
    \left(\frac{g_-^2}{g_+^2}\right)^{N_-}~,
\end{equation}
are sufficient to render every individual diagram finite. Thus, after
including the external $\langle JA\rangle$ propagators, there are no
separately divergent diagrams in the $--+\,+$, $-+-\,+$, $---\,+$, or $---\,-$
sectors.

For the alternating-helicity correlator, all finite contributions are proportional to $\Pi^+$. We suppress the common tensor
\begin{equation}
    \epsilon_i^-(\vec k_1)\epsilon_j^+(\vec k_2)
    \epsilon_k^-(\vec k_3)\epsilon_l^+(\vec k_4)
    \epsilon_{ijm}\epsilon_{kln}~.
\end{equation}

The finite Yang--Mills exchange contribution is
\begin{align}
\left.\langle J^-J^+J^-J^+\rangle_s
\right|_{\mathrm{YM-YM}}^{\mathrm{fin}}
=g_+^6
\frac{(k_2-k_1+q)(k_4-k_3+q)}
{4q E_LE_R}
\Pi^+_{mn}(\vec q)~.
\end{align}
The sum of the mixed diagrams gives
\begin{align}
\left.\langle J^-J^+J^-J^+\rangle_s
\right|_{\mathrm{mixed}}^{\mathrm{fin}}
=
-\frac{g_+^6}{4q}
\left[
\frac{k_2-k_1+q}{E_L}
+\frac{k_4-k_3+q}{E_R}
\right]
\Pi^+_{mn}(\vec q)~.
\end{align}
The Chern--Simons exchange contribution is
\begin{align}
\left.\langle J^-J^+J^-J^+\rangle_s
\right|_{\mathrm{CS-CS}}^{\mathrm{fin}}
=
\frac{g_+^6}{4q}\Pi^+_{mn}(\vec q)~,
\end{align}
whereas the quartic contact diagram vanishes in this limit:
\begin{equation}
\left.\langle J^-J^+J^-J^+\rangle_s
\right|_{\mathrm{contact}}^{\mathrm{fin}}
=0~.
\end{equation}
The sum factorizes as
\begin{align}
&\frac{g_+^6}{4q}
\left[
1-\frac{k_2-k_1+q}{E_L}
\right]
\left[
1-\frac{k_4-k_3+q}{E_R}
\right]
\Pi^+_{mn}(\vec q)
=g_+^6
\frac{k_1k_3}{qE_LE_R}
\Pi^+_{mn}(\vec q)~.
\end{align}
Restoring the polarizations and adding the $t$-channel gives the third result in \eqref{eq:4points}. For the adjacent-helicity ordering $\langle J^- J^- J^+ J^+  \rangle$, the $s$-channel vanishes,
while its $t$-channel is obtained from the alternating-helicity result
by the appropriate permutation, which reproduces the fourth result in \eqref{eq:4points}.

Finally, in the self-dual limit, every diagram vanishes in
the $\langle J^- J^- J^- J^+ \rangle$ and $\langle J^- J^- J^- J^- \rangle$ sectors. Thus, the sum of
the finite pieces of the second-order diagrams precisely reproduces all
the nonvanishing four-point correlators obtained directly in the
first-order formulation.

\section{Propagators in BF theory}
\label{sec:Propagators in BF theory}

The bulk-to-bulk propagators of the BF action \eqref{eq:BF_action} for Yang--Mills theory are obtained by inverting the kinetic term, which can be written as
\bea \frac{1}{2}
\begin{pmatrix}   A_i  &b_i \end{pmatrix}
\begin{pmatrix}
0
&
-2\delta_{ik}\partial_z-2\epsilon_{ijk}\partial_j
\\[0.8em]
2\delta_{ik}\partial_z-2\epsilon_{ijk}\partial_j
&
-4g^2\delta_{ik}
\end{pmatrix}  \begin{pmatrix}   A_k  \\ b_k \end{pmatrix}~.
\eea
In Fourier space at the AdS boundary, the bulk-to-bulk propagators satisfy
\bea
\label{eq:B2B}
&\begin{pmatrix}
0
&
-2\delta_{ik}\partial_z-2i\epsilon_{ijk}k_j
\\[0.8em]
2\delta_{ik}\partial_z-2i \epsilon_{ijk}k_j
&
-4g^2\delta_{ik}
\end{pmatrix}
\begin{pmatrix}
\left\langle A_k^a(z) A_\ell^b(z') \right\rangle
&
\left\langle A_k^a(z) b_\ell^b(z') \right\rangle
\\[0.8em]
\left\langle b_k^a(z) A_\ell^b(z') \right\rangle
&
\left\langle b_k^a(z) b_\ell^b(z') \right\rangle
\end{pmatrix}\\
&=
\delta^{ab}
\begin{pmatrix}
\delta_{i\ell} & 0
\\[0.4em]
0 & \delta_{i\ell}
\end{pmatrix}
\delta(z-z')~.
\eea 
The most general solution compatible with regularity in the interior is given by
\begin{align}
\left\langle A_i^a(\vec k, z) A_j^b(-\vec k, z') \right\rangle &=  g^2 \delta^{ab} \left[ G^L (k, z, z') L_{ij} + G^+ (k, z, z') \Pi^+_{ij}  +   G^- (k, z, z') \Pi^-_{ij}  \right] \, ,  \nonumber \\
\left\langle A_i^a(\vec k, z) b_j^b(-\vec k, z') \right\rangle
&=
\delta^{ab}
\Bigg[
\frac{1}{4}\left(\text{sign} (z-z')+s_L\right)L_{ij}
+
\left(
\frac{1}{2}\theta(z-z')\,e^{-k|z'-z|}
+s_+\,e^{-k(z+z')}
\right)\Pi^+_{ij} \nonumber
\\
&\qquad\qquad\qquad\qquad\qquad+
\left(
-\frac{1}{2}\theta(z'-z)\,e^{-k|z'-z|}
+s_-\,e^{-k(z+z')}
\right)\Pi^-_{ij}
\Bigg] \, , \nonumber
\\
\left\langle b_i^a(\vec k, z) A_j^b(-\vec k , z') \right\rangle
&=
\delta^{ab}
\Bigg[
-\frac{1}{4}\left(\text{sign} (z-z')+r_L\right)L_{ij}
+\frac{1}{2}\theta(z'-z)\,e^{-k|z'-z|}\,\Pi^+_{ij} \nonumber
\\
&\qquad\qquad\qquad\qquad\qquad-\frac{1}{2}
\left(
\theta(z-z')\,e^{-k|z'-z|}
+r_-\,e^{-k(z+z')}
\right)\Pi^-_{ij}
\Bigg] \, ,  \nonumber \\
\left\langle b_i^a(\vec k, z)b_j^b(-\vec k, z') \right\rangle
&=-
\frac{k}{g^2}\,
\delta^{ab}\,
s_-\,
e^{-k(z+z')}\,
\Pi^-_{ij}~, \label{eq:B2Bmostgeneral}
\end{align}  where 
\begin{equation} 
\begin{split}
    G^\pm (k, z, z') &= \frac{1}{2 k} (e^{-k |z'-z|} + r_{\pm}  e^{-k |z'+ z|}  ), \\
    G^L (k, z, z') &= -\frac{1}{2} (|z'-z| + r_L |z'+z| ) + l(k)~. 
\end{split}
\end{equation}
The projectors are defined in Equation~\eqref{projectors}. Furthermore, $r_\pm$, $r_L$, $s_\pm$, and $s_L$ are six functions of $k$ that we now fix using consistency and boundary conditions.

First, requiring the consistency condition
\begin{equation}
\left\langle A_i^a(\vec k,z)b_j^b(-\vec k,z')\right\rangle
=
\left\langle b_j^b(-\vec k,z')A_i^a(\vec k,z)\right\rangle~,
\end{equation} which is expected for a propagator, we find
\begin{equation}
    s_+ = 0, \qquad s_- = -\frac{r_-}{2}, \qquad s_L = - r_L~,
\end{equation} so that 
\begin{equation} \label{eq:B2Bgeneral}
\begin{split}
\left\langle A_i^a(\vec k, z) A_j^b(-\vec k , z') \right\rangle
&=
g^2 \delta^{ab}
\left[
G^L(k,z,z') L_{ij}
+
G^+(k,z,z') \Pi^+_{ij}
+
G^-(k,z,z') \Pi^-_{ij}
\right] \, ,
\\
\left\langle A_i^a(\vec k , z) b_j^b(-\vec k , z' ) \right\rangle
&=
\delta^{ab}
\Bigg[
\frac{1}{4}\left(\text{sign} (z-z')-r_L\right)L_{ij}
+
\frac{1}{2}\theta(z-z')\,e^{-k|z-z'|}\,\Pi^+_{ij}
\\
&\hspace{4.5em}
-\frac{1}{2}
\left(
\theta(z'-z)\,e^{-k|z-z'|}
+
r_-\,e^{-k(z+z')}
\right)\Pi^-_{ij}
\Bigg] \, ,
\\
\left\langle b_i^a(\vec k, z) A_j^b(-\vec k, z') \right\rangle
&=
\delta^{ab}
\Bigg[
-\frac{1}{4}\left(\text{sign} (z-z')+r_L\right)L_{ij}
+
\frac{1}{2}\theta(z'-z)\,e^{-k|z-z'|}\,\Pi^+_{ij}
\\
&\hspace{4.5em}
-\frac{1}{2}
\left(
\theta(z-z')\,e^{-k|z-z'|}
+
r_-\,e^{-k(z+z')}
\right)\Pi^-_{ij}
\Bigg] \, ,
\\
\left\langle b_i^a(\vec k, z)b_j^b(-\vec k , z') \right\rangle
&=
\frac{k}{2g^2}\,
\delta^{ab}\,
r_-\,e^{-k(z+z')}\,
\Pi^-_{ij}~.
\end{split}
\end{equation}

Finally, imposing the Neumann boundary conditions \eqref{eq:N_BF}, which, in axial gauge, reduces to
\begin{equation}
   \left( \frac{1}{g_+^2}\star F_{zi}+2B_{zi}=0 \right) \Big|_{z=0}~, 
\end{equation} implies the following constraints on the remaining undetermined functions: 
\be r_\pm = \bigg(\frac{g_+}{g_-}\bigg)^{\pm 2} \,, \qquad  r_L = +1~.
\ee 
Taking this condition into account, the propagators \eqref{eq:B2Bgeneral} reproduce \eqref{eq:B2Bafterboundarycond}.

\section{Four-point functions in the BF formulation}
\label{sec:Four-point functions in the BF formulation}

In this appendix, we derive the four-point correlators directly from
the Witten diagrams of the first-order BF formulation. We use the self-dual bulk-to-bulk \eqref{eq:B2Bsd} and bulk-to-boundary \eqref{eq:B2b} propagators, as well as the BF and Chern--Simons vertices; see Figure~\ref{fig:4pt}.


\subsection{$\langle J^+J^+J^+J^+\rangle$}

For the all-plus correlator, we suppress the common polarization tensor
in the intermediate expressions:
\begin{equation}
    \epsilon_i^+(\vec k_1)
    \epsilon_j^+(\vec k_2)
    \epsilon_k^+(\vec k_3)
    \epsilon_l^+(\vec k_4)
    \epsilon_{ijm}\epsilon_{kln}~.
\end{equation}
There are three types of $s$-channel exchange diagrams. First, the
diagram containing two boundary Chern--Simons vertices is connected
by an $AA$ propagator and gives
\begin{equation}
\left.
\left\langle J^+J^+J^+J^+\right\rangle_s
\right|_{\mathrm{CS-CS}}
=
g_+^6
\frac{\Pi^+_{mn}(\vec q)}{q}~.
\label{eq:BF-pppp-CSCS}
\end{equation}
The diagram containing two bulk BF vertices is connected by a $bb$
propagator. Performing the two radial integrals gives
\begin{equation}
\left.
\left\langle J^+J^+J^+J^+\right\rangle_s
\right|_{\mathrm{BF-BF}}
=
g_+^6
\frac{q}{E_LE_R}
\Pi^-_{mn}(\vec q)~.
\label{eq:BF-pppp-BFBF}
\end{equation}
Finally, there are two mixed diagrams containing one BF  vertex and one
Chern--Simons vertex. Using the mixed propagator with one point at the
boundary, their sum is
\begin{align}
\left.
\left\langle J^+J^+J^+J^+\right\rangle_s
\right|_{\mathrm{mixed}}
=
-g_+^6
\Bigg[
&\left(
\frac{1}{E_L}
+
\frac{1}{E_R}
\right)
\Pi^-_{mn}(\vec q)+
\left(
\frac{1}{k_{12}}
+
\frac{1}{k_{34}}
\right)
L_{mn}(\vec q)
\Bigg]~.
\label{eq:BF-pppp-mixed}
\end{align}
Adding the three contributions and using
\begin{equation}
    E_L+E_R=E+2q,
    \qquad
    k_{12}+k_{34}=E~,
\end{equation}
we obtain
\begin{align}
\left\langle J^+J^+J^+J^+\right\rangle_s
=
g_+^6
\left[
\frac{\Pi^+_{mn}(\vec q)}{q}
-
\frac{E+q}{E_LE_R}\Pi^-_{mn}(\vec q)
-
\frac{E}{k_{12}k_{34}}L_{mn}(\vec q)
\right]~.
\label{eq:BF-pppp-sum}
\end{align}
Restoring the polarization tensor and including the $t$-channel gives
\begin{align}
&\left\langle
J^+(\vec k_1)J^+(\vec k_2)
J^+(\vec k_3)J^+(\vec k_4)
\right\rangle_{\mathrm{SD}}
\nonumber\\
&=
g_+^6
\epsilon_i^+(\vec k_1)
\epsilon_j^+(\vec k_2)
\epsilon_k^+(\vec k_3)
\epsilon_l^+(\vec k_4)
\epsilon_{ijm}\epsilon_{kln}
\nonumber\\
&\quad\times
\left[
\frac{\Pi^+_{mn}(\vec q)}{q}
-
\frac{E+q}{E_LE_R}\Pi^-_{mn}(\vec q)
-
\frac{E}{k_{12}k_{34}}L_{mn}(\vec q)
\right]
+
(2\leftrightarrow4)~.
\end{align}

\subsection{$\langle J^-J^+J^+J^+\rangle$}

For the one-minus correlator, we suppress the common polarization
tensor
\begin{equation}
    \epsilon_i^-(\vec k_1)
    \epsilon_j^+(\vec k_2)
    \epsilon_k^+(\vec k_3)
    \epsilon_l^+(\vec k_4)
    \epsilon_{ijm}\epsilon_{kln}~.
\end{equation}
Since a negative-helicity current couples to $b$, the vertex containing
the external legs $1$ and $2$ must be a BF vertex. The diagram with a
BF vertex on the left and a Chern--Simons vertex on the right is
connected by an $AA$ propagator and gives
\begin{equation}
\left.
\left\langle J^-J^+J^+J^+\right\rangle_s
\right|_{\mathrm{BF-CS}}
=
g_+^6 k_1
\frac{\Pi^+_{mn}(\vec q)}{qE_L}~.
\label{eq:BF-mppp-BFCS}
\end{equation}
The diagram containing two  BF vertices is connected by the mixed
$Ab$ propagator in \eqref{eq:B2Bsd}. Using the
ordered integrals \eqref{eq:BF-ordered-integrals}, we find
\begin{align}
\left.
\left\langle J^-J^+J^+J^+\right\rangle_s
\right|_{\mathrm{BF-BF}}
=
g_+^6 k_1
\left[
\frac{\Pi^+_{mn}(\vec q)}{EE_L}
-
\frac{\Pi^-_{mn}(\vec q)}{EE_R}
-
\frac{L_{mn}(\vec q)}{Ek_{34}}
\right]~.
\label{eq:BF-mppp-BFBF}
\end{align}
Their sum is therefore
\begin{align}
\left\langle J^-J^+J^+J^+\right\rangle_s
=
g_+^6 k_1
\left[
\frac{\Pi^+_{mn}(\vec q)}{E_L}
\left(
\frac{1}{q}
+
\frac{1}{E}
\right)
-
\frac{\Pi^-_{mn}(\vec q)}{EE_R}
-
\frac{L_{mn}(\vec q)}{Ek_{34}}
\right]~.
\label{eq:BF-mppp-sum}
\end{align}
Restoring the polarization tensor and adding the $t$-channel gives
\begin{align}
&\left\langle
J^-(\vec k_1)J^+(\vec k_2)
J^+(\vec k_3)J^+(\vec k_4)
\right\rangle_{\mathrm{SD}}
\nonumber\\
&=
g_+^6 k_1
\epsilon_i^-(\vec k_1)
\epsilon_j^+(\vec k_2)
\epsilon_k^+(\vec k_3)
\epsilon_l^+(\vec k_4)
\epsilon_{ijm}\epsilon_{kln}
\nonumber\\
&\quad\times
\left[
\frac{\Pi^+_{mn}(\vec q)}{E_L}
\left(
\frac{1}{q}
+
\frac{1}{E}
\right)
-
\frac{\Pi^-_{mn}(\vec q)}{EE_R}
-
\frac{L_{mn}(\vec q)}{Ek_{34}}
\right]
+
(2\leftrightarrow4)~.
\end{align}

\subsection{$\langle J^-J^+J^-J^+\rangle$}

For the alternating-helicity correlator, we suppress the common
polarization tensor
\begin{equation}
    \epsilon_i^-(\vec k_1)
    \epsilon_j^+(\vec k_2)
    \epsilon_k^-(\vec k_3)
    \epsilon_l^+(\vec k_4)
    \epsilon_{ijm}\epsilon_{kln}~.
\end{equation}
Each pair of external legs contains one negative-helicity current.
Consequently, both vertices must be $BAA$ vertices, and the internal line
is an $AA$ propagator. The $s$-channel contribution is
\begin{equation}
\left\langle J^-J^+J^-J^+\right\rangle_s
=
g_+^6 k_1k_3
\frac{\Pi^+_{mn}(\vec q)}
{qE_LE_R}~.
\label{eq:BF-mpmp-s}
\end{equation}
Restoring the polarization tensor and adding the $t$-channel gives
\begin{align}
&\left\langle
J^-(\vec k_1)J^+(\vec k_2)
J^-(\vec k_3)J^+(\vec k_4)
\right\rangle_{\mathrm{SD}}
\nonumber\\
&=
g_+^6 k_1 k_3
\epsilon_i^-(\vec k_1)
\epsilon_j^+(\vec k_2)
\epsilon_k^-(\vec k_3)
\epsilon_l^+(\vec k_4)
\epsilon_{ijm}\epsilon_{kln}
\frac{\Pi^+_{mn}(\vec q)}
{qE_LE_R}
+
(2\leftrightarrow4)~.
\end{align}

\subsection{Adjacent and mostly-minus helicity configurations}

A convenient cyclic representative of the configuration with two
adjacent negative-helicity insertions is
$\langle J^-J^+J^+J^-\rangle$. In its $s$-channel, each  vertex
contains one external $b$ leg, and the two BF vertices are connected by
an $AA$ propagator. We obtain
\begin{align}
&\left\langle
J^-(\vec k_1)J^+(\vec k_2)
J^+(\vec k_3)J^-(\vec k_4)
\right\rangle_{\mathrm{SD}}
\nonumber\\
&=g_+^6
k_1 k_4
\epsilon_i^-(\vec k_1)
\epsilon_j^+(\vec k_2)
\epsilon_k^+(\vec k_3)
\epsilon_l^-(\vec k_4)
\epsilon_{ijm}\epsilon_{kln}
\frac{\Pi^+_{mn}(\vec q)}
{qE_LE_R}~.
\label{eq:BF-adjacent}
\end{align}
The other planar channel vanishes because its two negative-helicity
external legs would meet at the same  vertex, whereas the $BAA$
interaction contains only one $b$ field. By cyclicity, this result also
determines the correlator
$\langle J^-J^-J^+J^+\rangle$ after the corresponding relabeling of
the external momenta.

Finally, any four-point exchange diagram contains at most two 
vertices and therefore at most two external $b$ legs. It follows that
\begin{equation}
\left\langle J^-J^-J^-J^+\right\rangle_{\mathrm{SD}}
=
\left\langle J^-J^-J^-J^-\right\rangle_{\mathrm{SD}}
=0~.
\end{equation}
Thus, the Witten-diagram computation in the first-order 
formulation reproduces all the finite four-point correlators obtained
by taking the self-dual limit of the second-order formulation.

\section{Correlators of the non-linear operator}
\label{sec:Correlators of the non-linear operator}
In this appendix, we compute the tree-level connected part of correlators of the full boundary operator
\begin{equation}
    \mathcal J_i^a(\vec k)=-\frac{1}{2} \epsilon_{ijk} F_{jk}^a(\vec{k}, 0)=J_i^a(\vec k)+Q_i^a(\vec k) \;,
\label{eq:full-current-split}
\end{equation}
which we distinguish from its linear part
\begin{equation}
    J_i^a(\vec k)=-i\epsilon_{ijk}k_jA_k^a(\vec k,0) \;,
\end{equation}
whose correlators are calculated in Sections \ref{sec:Self-dual holographic correlators} and \ref{sec:Self-dual holographic correlators2}. 
The non-linear part is given by
\begin{equation}
    Q_i^a(\vec k)
    =-\frac{1}{2}f^{abc}\epsilon_{ijk}
      \int\frac{d^3\vec p}{(2\pi)^3}
      A_j^b(\vec p,0)A_k^c(\vec k-\vec p,0).
\label{eq:quadratic-current-vertex}
\end{equation}
Below, we compute the correlators of the full $\mathcal J_i^a$, which is the 3d Hodge dual of the boundary field strength of $A_i^a(\vec{k},0)$. $Q$ is a boundary composite operator, whose self-contractions do not contribute to tree-level diagrams.

Note that for multiple $Q$ insertions to contribute to the tree-level correlator, at least one of them, $Q(\vec{k}_i)$, must Wick contract with one of the external $J(\vec{k}_j)$. Then, if there is a bulk interaction vertex $\int_0^\infty dz$ which connects to the other $A$ in $Q$, the incoming momentum to that vertex will be $\vec{k}_i+\vec{k}_j$, which means that the energy flowing into that vertex will come from $e^{-|\vec{k}_i+\vec{k}_j|z}$. In other words, this contribution can only depend on $|\vec{k}_i+\vec{k}_j|$ but not on $|\vec{k}_i| + |\vec{k}_j|$, which means that the radial integrals of the vertex can not produce a total energy pole in this diagram. Hence, the non-linear corrections due to $Q$ computed in this appendix can not change the total energy poles computed from the correlators of the linearized operator $J$, and hence do not alter the flat-space amplitudes extracted from them.

\subsection{Yang--Mills plus $\theta$-term correlators}
\label{app:nonlinear-current-second-order}
First, we compute correlators of $\mathcal{J}_i^a$ in the second-order formulation. At the self-dual point, $Q$ does not contribute to the two-point function at tree level. At three points, having a single $Q$ Wick contracted with the other two $J$'s is a tree-level diagram of the same order in $g_+$ as the $\langle JJJ\rangle$ cubic vertex contribution. At four points, having a single $Q$ with a cubic vertex and having two $Q$'s Wick-contracted with the other $J$'s are of the same order as the $\langle JJJJ\rangle$ correlator. Having more $Q$'s is subleading in $g_+$ and corresponds to loop corrections.

It is useful to record the two free boundary contractions
\begin{align}
 \left\langle J_i^a(\vec k)A_j^b(-\vec k,0)\right\rangle
 &=\delta^{ab}P_{ij}(\vec k),
 &
 P_{ij}(\vec k)
 &=g_+^2\Pi^+_{ij}(\vec k)-g_-^2\Pi^-_{ij}(\vec k),
\label{eq:PA-def}\\
 \left\langle A_i^a(\vec k,0)A_j^b(-\vec k,0)\right\rangle
 &=\delta^{ab}G_{ij}(\vec k),
 &
 G_{ij}(\vec k)
 &=\frac{g_+^2}{k}\Pi^+_{ij}(\vec k)
   +\frac{g_-^2}{k}\Pi^-_{ij}(\vec k)
   +2\frac{g_+^2g_-^2}{g_+^2+g_-^2}l(k)L_{ij}(\vec k).
\label{eq:AA-boundary-def}
\end{align}
The last term keeps track of the residual boundary gauge choice, which still drops out in the self-dual limit considered below.

\subsubsection{Two-point function}

There is no tree-level correction to the connected two-point function:
\begin{equation}
 \left\langle\mathcal J_i^a(\vec k)
 \mathcal J_j^b(-\vec k)\right\rangle
 =
 \left\langle J_i^a(\vec k)J_j^b(-\vec k)\right\rangle.
\label{eq:full-current-two-point-general}
\end{equation}
Indeed, $\langle QJ\rangle$ contains an odd number of gauge fields,
whereas the connected Gaussian contraction $\langle QQ\rangle$ is a
one-loop boundary bubble.  Hence
\begin{equation}
 \left\langle\mathcal J_i^a(\vec k)
 \mathcal J_j^b(-\vec k)\right\rangle
 =\delta^{ab}k
 \left[g_+^2\Pi^+_{ij}(\vec k)+g_-^2\Pi^-_{ij}(\vec k)\right]
\label{eq:full-current-two-point-finite-couplings}
\end{equation} and the self-dual limit yields \eqref{2ptsdfull}. 

\subsubsection{Three-point function}

At three points the quadratic part of one insertion contributes
at tree level.  Wick's theorem gives
\begin{align}
&\Delta_Q
\left\langle
\mathcal J_{i}^{a}(\vec k_1)
\mathcal J_{j}^{b}(\vec k_2)
\mathcal J_{k}^{c}(\vec k_3)
\right\rangle
\nonumber\\
&\quad=-f^{abc}
\Big[
 \epsilon_{imn}P_{jm}(\vec k_2)P_{kn}(\vec k_3)
 +\epsilon_{jmn}P_{km}(\vec k_3)P_{in}(\vec k_1)
 +\epsilon_{kmn}P_{im}(\vec k_1)P_{jn}(\vec k_2)
\Big] \;.
\label{eq:nonlinear-three-point-tensor}
\end{align}
The factor $1/2$ in \eqref{eq:quadratic-current-vertex} is cancelled by
the two Wick contractions.  This contribution contains no bulk radial
integral and is regular at the total-energy pole $E=0$. Upon contraction with the external polarization vectors with helicity $\sigma_i$, we get
\begin{equation}
    \Delta_Q \left\langle \mathcal{J}^{\sigma_1,a}(\vec{k}_1) \mathcal{J}^{\sigma_2,b}(\vec{k}_2) \mathcal{J}^{\sigma_3,c}(\vec{k}_3)\right \rangle = -f^{abc}\epsilon_{ijk}
 \epsilon_i^{\sigma_1}(\vec k_1)\epsilon_j^{\sigma_2}(\vec k_2)
 \epsilon_k^{\sigma_3}(\vec k_3) [\sigma_1 \sigma_2 g_{\sigma_1}^2 g_{\sigma_2}^2 +\text{cyclic}] ~.
\end{equation}

After stripping the color factor, the full finite-$g_-$
correlators in the helicity basis are
\begin{align}
\left\langle\mathcal J^+(\vec k_1)
\mathcal J^+(\vec k_2)\mathcal J^+(\vec k_3)\right\rangle
&=-2g_+^4\epsilon_{ijk}
 \epsilon_i^+(\vec k_1)\epsilon_j^+(\vec k_2)
 \epsilon_k^+(\vec k_3),
\nonumber\\
\left\langle\mathcal J^-(\vec k_1)
\mathcal J^+(\vec k_2)\mathcal J^+(\vec k_3)\right\rangle
&=\left[
 \frac{g_+^2\bigl(g_+^2k_1-g_-^2(k_2+k_3)\bigr)}{E}
 -g_+^4+2g_+^2g_-^2
 \right]
 \epsilon_{ijk}\epsilon_i^-(\vec k_1)
 \epsilon_j^+(\vec k_2)\epsilon_k^+(\vec k_3),
\nonumber\\
\left\langle\mathcal J^-(\vec k_1)
\mathcal J^-(\vec k_2)\mathcal J^+(\vec k_3)\right\rangle
&=\left[
 \frac{g_-^2\bigl(g_-^2k_3-g_+^2(k_1+k_2)\bigr)}{E}
 -g_-^4+2g_+^2g_-^2
 \right]
 \epsilon_{ijk}\epsilon_i^-(\vec k_1)
 \epsilon_j^-(\vec k_2)\epsilon_k^+(\vec k_3),
\nonumber\\
\left\langle\mathcal J^-(\vec k_1)
\mathcal J^-(\vec k_2)\mathcal J^-(\vec k_3)\right\rangle
&=-2g_-^4\epsilon_{ijk}
 \epsilon_i^-(\vec k_1)\epsilon_j^-(\vec k_2)
 \epsilon_k^-(\vec k_3).
\label{eq:full-current-three-point-finite-couplings}
\end{align}
These expressions are the sum of \eqref{eq:nonlinear-three-point-tensor}
and the Yang--Mills plus Chern--Simons diagrams computed with the
linearized operator. Importantly, they are finite as $g_- \to 0$, and the self-dual correlators reproduce \eqref{eq:full-current-three-point-SD}, which have the same total energy poles as \eqref{3pt correlators}. 

\subsubsection{Four-point function}

At four points the complete nonlinear correction contains only two
types of tree diagrams.  The first contains one $Q$ insertion and one
ordinary cubic interaction vertex; the second contains two $Q$
insertions and no interaction vertex. In particular, we have
\begin{equation}
 \left\langle\mathcal J_1\mathcal J_2
 \mathcal J_3\mathcal J_4\right\rangle
 =
 \left\langle J_1J_2J_3J_4\right\rangle
 +\Delta_{1Q}+\Delta_{2Q},
\label{eq:four-point-full-Q-expansion}
\end{equation}
where $J_r\equiv J_{i_r}^{a_r}(\vec k_r)$, and the color-ordered non-linear correction is given 
\begin{equation} \label{eq:Delta1Q_4_color-ordered}
    \begin{split}
        \Delta_{1Q} &= \quad  \left[ P_{i_1m}(\vec{k}_1)\epsilon_{i_2 m n} - P_{i_2 m} (\vec k_2)\epsilon_{i_1 mn} \right] \left\langle A_n(\vec k_1 + \vec k_2,0) J_{i_3}(\vec k_3 ) J_{i_4}(\vec k_4)  \right\rangle + (2 \leftrightarrow 4) \\
        &\quad +\left[ P_{i_3m}(\vec k_3) \epsilon_{i_4 mn} - P_{i_4 m}(\vec k_4) \epsilon_{i_3 mn} \right] \left\langle A_n(\vec k_3 + \vec k_4,0) J_{i_1}(\vec k_1 ) J_{i_2}(\vec k_2)  \right\rangle +  (2 \leftrightarrow 4) \;,
    \end{split}
\end{equation}
and
\begin{equation} \label{eq:Delta2Q_4_color-ordered}
    \Delta_{2 Q} = \left[ P_{i_1m}(\vec{k}_1)\epsilon_{i_2 m n} - P_{i_2 m} (\vec k_2)\epsilon_{i_1 mn} \right] G_{nl}(\vec{k}_1 + \vec k_2) \left[ P_{i_3p}(\vec k_3) \epsilon_{i_4 p l} - P_{i_4 p}(\vec k_4) \epsilon_{i_3 pl} \right] +(2\leftrightarrow4) \;.
\end{equation}
The purely Wick-contracted contribution
$\Delta_{2Q}$ has a smooth $g_- \to 0$ limit. On the other hand, the three-point function that contributes to $\Delta_{1Q}$ is computed with
the linear operator $J$ and is the sum of the bulk Yang--Mills and
boundary Chern--Simons diagrams,
\begin{align}
    \left\langle
    A_i^a(\vec{p}_1,0)
    J_j^b(\vec{p}_2)
    J_k^c(\vec{p}_3)
    \right\rangle&=
    f^{abc}
    P_{jm}(\vec{p}_2)P_{kn}(\vec{p}_3) \nonumber \\
    &\Bigg\{
    -\frac{i}{2}
    \left(
    \frac{1}{g_+^2}+\frac{1}{g_-^2}
    \right)
    \Big[
    \delta_{\ell m}(\vec{p}_1-\vec{p}_2)_n
    +\delta_{mn}(\vec{p}_2-\vec{p}_3)_\ell
    +\delta_{n\ell}(\vec{p}_3-\vec{p}_1)_m
    \Big]
    \nonumber\\
    &\qquad\times
    \Bigg[
    \frac{
    g_+^2\Pi^+_{i\ell}(\vec{p}_1)
    +g_-^2\Pi^-_{i\ell}(\vec{p}_1)}
    {p_1(p_1+p_2+p_3)}
    +
    \frac{2g_+^2g_-^2}{g_+^2+g_-^2}
    \left(
    \frac{l(p_1)}{p_2+p_3}
    -\frac{1}{(p_2+p_3)^2}
    \right)
    L_{i\ell}(\vec{p}_1)
    \Bigg]
    \nonumber\\
    &\quad
    +\frac{1}{2}
    \left(
    \frac{1}{g_+^2}-\frac{1}{g_-^2}
    \right)
    \epsilon_{\ell mn}
    G_{i\ell}(\vec{p}_1)
    \Bigg\}.
\end{align}
The two diagrams diverge separately as $g_- \to 0$, but the divergences cancel after summing them, just as for the linearized-operator diagrams, and we get 
\begin{equation}
\label{eq:AJJ_SD}
    \begin{split}
        \left\langle A_i^a(\vec{p}_1) J_j^b(\vec p_2)  J^c_k(\vec p_3) \right\rangle \xrightarrow{g_-=0} f^{abc} g_+^4 \epsilon_{lmn} \Bigg[& \frac{\Pi_{il}^+(\vec{p}_1) \Pi^+_{jm}(\vec p_2) \Pi^+_{kn}(\vec p_3)}{p_1} \\
        &- \frac{\Pi_{il}^-(\vec{p}_1) \Pi^+_{jm}(\vec p_2) \Pi^+_{kn}(\vec p_3)}{p_1+p_2+p_3} -\frac{L_{il}(\vec p_1) \Pi^+_{jm}(\vec p_2) \Pi^+_{kn}(\vec p_3)}{p_2+p_3} \\
        &+ \frac{p_2 \Pi_{il}^+(\vec{p}_1) \Pi^-_{jm}(\vec p_2) \Pi^+_{kn}(\vec p_3)}{p_1(p_1+p_2+p_3)} +\frac{p_3 \Pi_{il}^+(\vec{p}_1) \Pi^+_{jm}(\vec p_2) \Pi^-_{kn}(\vec p_3)}{p_1(p_1+p_2+p_3)} \Bigg] \;.
    \end{split}
\end{equation}
Thus the complete tree correlators
are finite in the self-dual limit, although not every second-order diagram is separately finite.

The nonlinear contributions in the self-dual
limit are displayed explicitly below.  We give the color-ordered
$s$-channel; the $t$-channel is obtained as in the main text.  For the
all-plus sector,
\begin{align}
\left.\Delta_Q\langle
\mathcal J^+(\vec k_1)\mathcal J^+(\vec k_2)
\mathcal J^+(\vec k_3)\mathcal J^+(\vec k_4)
\rangle_s\right|_{\rm sd}
={}&2g_+^6
\epsilon_i^+(\vec k_1)\epsilon_j^+(\vec k_2)
\epsilon_k^+(\vec k_3)\epsilon_l^+(\vec k_4)
\epsilon_{ijm}\epsilon_{kln}
\nonumber\\[-0.2em]
&\times\left[
\frac{E+2q}{E_LE_R}\Pi^-_{mn}(\vec q)
+\frac{E}{k_{12}k_{34}}L_{mn}(\vec q)
\right].
\label{eq:Q-correction-pppp-sd}
\end{align}
For one negative helicity,
\begin{align}
\left.\Delta_Q\langle
\mathcal J^-(\vec k_1)\mathcal J^+(\vec k_2)
\mathcal J^+(\vec k_3)\mathcal J^+(\vec k_4)
\rangle_s\right|_{\rm sd}
={}&g_+^6
\epsilon_i^-(\vec k_1)\epsilon_j^+(\vec k_2)
\epsilon_k^+(\vec k_3)\epsilon_l^+(\vec k_4)
\epsilon_{ijm}\epsilon_{kln}
\nonumber\\[-0.2em]
&\times\left[
\frac{k_2+q-k_1}{qE_L}\Pi^+_{mn}(\vec q)
+\frac{1}{E_R}\Pi^-_{mn}(\vec q)
+\frac{1}{k_{34}}L_{mn}(\vec q)
\right].
\label{eq:Q-correction-mppp-sd}
\end{align}
For the alternating and adjacent two-minus orderings,
\begin{align}
\left.\Delta_Q\langle
\mathcal J^-(\vec k_1)\mathcal J^+(\vec k_2)
\mathcal J^-(\vec k_3)\mathcal J^+(\vec k_4)
\rangle_s\right|_{\rm sd}
={}&\frac{g_+^6}{q}
\epsilon_i^-(\vec k_1)\epsilon_j^+(\vec k_2)
\epsilon_k^-(\vec k_3)\epsilon_l^+(\vec k_4)
\epsilon_{ijm}\epsilon_{kln}
\nonumber\\[-0.2em]
&\times\left(1-\frac{k_1}{E_L}-\frac{k_3}{E_R}\right)
\Pi^+_{mn}(\vec q),
\label{eq:Q-correction-mpmp-sd}\\
\left.\Delta_Q\langle
\mathcal J^-(\vec k_1)\mathcal J^+(\vec k_2)
\mathcal J^+(\vec k_3)\mathcal J^-(\vec k_4)
\rangle_s\right|_{\rm sd}
={}&\frac{g_+^6}{q}
\epsilon_i^-(\vec k_1)\epsilon_j^+(\vec k_2)
\epsilon_k^+(\vec k_3)\epsilon_l^-(\vec k_4)
\epsilon_{ijm}\epsilon_{kln}
\nonumber\\[-0.2em]
&\times\left(1-\frac{k_1}{E_L}-\frac{k_4}{E_R}\right)
\Pi^+_{mn}(\vec q).
\label{eq:Q-correction-mppm-sd}
\end{align}
The nonlinear correction vanishes in the three-minus and all-minus
sectors.

Adding these terms to the correlators of the linearized operator gives the full four-point correlators at the self-dual point displayed in \eqref{4ptsdfull}. The other two-minus orderings follow by cyclicity.  Notice that the
nonlinear corrections do not alter the residues of the total-energy
poles; they only change terms regular at $E=0$.

\subsection{BF correlators}
\label{app:nonlinear-current-BF}

We now use the same operator definition
\eqref{eq:full-current-split}--\eqref{eq:quadratic-current-vertex} in the first-order formulation. At finite $g_\pm$, the $AA$ and $JA$ propagators agree with \eqref{eq:PA-def}--\eqref{eq:AA-boundary-def}. Since the non-linear corrections in the  formulation only come from Wick contractions of $A$, they exactly match the non-linear corrections in the second-order formulation computed above. Below, we compute  correlators using the full $\mathcal{J}$ at the self-dual point and verify this.

\subsubsection{Two- and three-point functions}

The two-point argument is unchanged, so there is no nonlinear
tree-level correction.  

At the self-dual point,
\begin{equation}
 \left\langle J_i^a(\vec k)A_j^b(-\vec k,0)\right\rangle
 =\delta^{ab}g_+^2\Pi^+_{ij}(\vec k).
\end{equation}
Equation \eqref{eq:nonlinear-three-point-tensor} therefore gives
\begin{align}
 \Delta_Q\langle\mathcal J^+\mathcal J^+\mathcal J^+\rangle
 &=-3g_+^4\epsilon_{ijk}\epsilon_i^+
   \epsilon_j^+\epsilon_k^+,
 &
 \Delta_Q\langle\mathcal J^-\mathcal J^+\mathcal J^+\rangle
 &=-g_+^4\epsilon_{ijk}\epsilon_i^-
   \epsilon_j^+\epsilon_k^+~.
\label{eq:BF-Q-three-point}
\end{align}
The correction vanishes when two or three external helicities are negative.
Adding the $BAA$ and Chern--Simons contact diagrams reproduces
\eqref{eq:full-current-three-point-SD}.

\subsubsection{Four-point function}

All  propagators, ordinary vertices, and composite-operator vertices
have smooth limits as $g_-\to0$ at fixed $g_+$.  Therefore every
tree-level  diagram contributing to the two-, three-, and four-point
functions above is individually finite, and we can compute the non-linear corrections exactly at the self-dual point. 

The non-linear corrections are given by $\Delta_{1Q}+\Delta_{2Q}$, which are given by the same expressions in \eqref{eq:Delta1Q_4_color-ordered}-\eqref{eq:Delta2Q_4_color-ordered}, since they involve Wick contractions of $A$ contained in $Q$ with the external legs $J_i \propto \epsilon_{ijk} k_j A_k$. The three-point function that contributes to $\Delta_{1Q}$ is computed at the self-dual point in \eqref{eq:AJJ_SD}. In the context of BF theory, the first term in \eqref{eq:AJJ_SD} arises from the Chern-Simons vertex, whereas the rest of the expression comes from the $BAA$ vertex. Computing the non-linear correction in each of the fixed helicity sectors gives \eqref{eq:Q-correction-pppp-sd}-\eqref{eq:Q-correction-mppm-sd}, with the final answer reproducing \eqref{4ptsdfull}.

\subsection{All multiplicity self-dual correlator} \label{app:n_pt correlator non lin}
In this section, we compute the analog of \eqref{eq:n pt_correlator} using the full non-linear operator. In particular, we compute the tree-level color-ordered $n$-point correlator
\begin{equation}
    \left\langle \mathcal J^+(\vec{k}_1) \mathcal J^-(\vec{k}_2) \dots \mathcal J^-(\vec{k}_{n-1}) \mathcal J^+(\vec{k}_n) \right\rangle \;
\end{equation}
at arbitrary multiplicity $n$. As discussed above, in order for the non-linear corrections to contribute to the tree-level correlator, one of the non-linear $Q$ insertions must Wick contract with one of the external legs, and we can have at most $n-2$ $Q$ insertions. 

It turns out that the non-linear correction to the $+$ external legs vanishes. This can be argued as follows. Every linear $J^-$ must connect to a $b$ leg (since $\langle J^-A\rangle = 0$), so it needs a  vertex. If we insert $Q^+$ instead of $J^+$, then we will need to form a loop in order to saturate the $A$ legs of the required  vertex. 

Hence, we only need to consider the non-linear corrections to the $-$ external legs. Since we have $n-2$ of them, which is the maximum allowed number of non-linear $Q$ insertions, the full answer can be computed as
\begin{equation}
\begin{split}
    &\left\langle \mathcal J^+(\vec{k}_1) \mathcal J^-(\vec{k}_2) \dots \mathcal J^-(\vec{k}_{n-1}) \mathcal J^+(\vec{k}_n) \right\rangle \\
    &\qquad= \left\langle \mathcal J^+(\vec{k}_1) \big(J^-(\vec{k}_2) + Q^-(\vec{k}_2) \big) \dots \big(J^-(\vec{k}_{n-1})+Q^-(\vec{k}_{n-1}) \big) \mathcal J^+(\vec{k}_n) \right\rangle \;.
\end{split}
\end{equation}
As shown in Section \ref{sec:n pt correlator}, $J^-(\vec{k}_j)$ connects to a  vertex and contributes a factor of 
\begin{equation}
    \frac{k_j}{k_j + \left| \sum_{i=1}^{j-1} \vec k_i \right|+ \left| \sum_{i=1}^{j} \vec k_i \right|} \;.
\end{equation}
In the color-ordered correlator, the corresponding $Q^-(\vec{k}_j)$ insertion replaces this factor with $-1$. Hence, the full non-linear $n$-point correlator can be computed by replacing
\begin{equation}
    \frac{k_j}{k_j + \left| \sum_{i=1}^{j-1} \vec k_i \right|+ \left| \sum_{i=1}^{j} \vec k_i \right|} \to \frac{k_j}{k_j + \left| \sum_{i=1}^{j-1} \vec k_i \right|+ \left| \sum_{i=1}^{j} \vec k_i \right|}-1 \;,
\end{equation}
which gives
\begin{equation} \label{eq:n pt_correlator non_lin}
    \begin{split}
        \bigg\langle \mathcal J^+(\vec{k}_1) \mathcal J^-(\vec{k}_2) & \dots \mathcal J^-(\vec{k}_{n-1}) \mathcal J^+(\vec{k}_n) \bigg\rangle \\
        &= g_+^{2n-2} \epsilon_{i_1}^+(\vec{k}_1) \epsilon_{i_2}^-(\vec{k}_2) \dots \epsilon_{i_{n-1}}^-(\vec{k}_{n-1}) \epsilon^+_{i_n} (\vec{k_n})  
        \\
        &\quad 
        \times  \epsilon_{i_1 i_2 j_2} \Pi^+_{j_2 k_2} (\vec{k}_1+\vec{k}_2)  \epsilon_{k_2 i_3 j_3} \Pi^+_{j_3 k_3}(\vec{k}_1 + \vec k_2+\vec k_3) \dots \Pi^+_{j_{n-2} k_{n-2}}(\vec k_1 + \dots + \vec k_{n-2}) \epsilon_{k_{n-2} i_{n-1} i_n} \\
        &\quad \times \left( \prod_{j=2}^{n-2} \frac{1}{\left| \sum_{i=1}^j \vec k_i \right|} \right) \cdot \prod_{j=2}^{n-1} \left( \frac{k_j}{k_j + \left| \sum_{i=1}^{j-1} \vec k_i \right|+ \left| \sum_{i=1}^{j} \vec k_i \right|} - 1 \right) \;.
    \end{split}
\end{equation}

\subsection{Alternative version of the full operator}
At the self-dual point, we have
\begin{equation}
    \star F_{zi}^a = F_{zi}^a ~.
\end{equation}
This means that the full boundary operator $\mathcal{J}_i^a$ at the self-dual point is equivalently given by $-F_{zi}$, which does not receive non-linear corrections in radial gauge $A_z=0$. Hence, self-dual holographic correlators could be equivalently computed using 
\begin{equation}
    \mathcal{J}_i^a = -F_{zi}^a = -\partial_z A_i^a \;.
\end{equation}

There will be no contact term differences between $F_{zi}$ and $\star F_{zi}$ correlators, which can be seen most clearly in the BF formulation where $F^-=0$ is the equation of motion of $B$:
\begin{equation}
    \langle F^-_{zi_1} F_{zi_2} \dots F_{zi_n} \rangle \propto \left \langle \frac{\delta}{\delta B_{zi_1}} ( F_{zi_2} \dots F_{zi_n} )\right\rangle = 0 \implies \langle \star F_{zi_1} F_{zi_2} \dots F_{zi_n} \rangle = \langle F_{zi_1} F_{zi_2} \dots F_{zi_n} \rangle \;.
\end{equation}
The bulk-to-boundary propagators computed using $F_{zi}$ have extra terms localized at $z=0$ relative to those computed using $\partial_{[i} A_{j]}$, which exactly match the non-linear corrections to $F_{jk}$ that we computed above.

\section{Spinor-helicity variables and polarization vectors}
\label{App:spinorHelicity}

We use the complex null lift of a boundary momentum employed in the main text,
\begin{equation}
 k^\mu=(i k,\vec{k}),\qquad k=|\vec{k}|,\qquad k^\mu k_\mu=0~,
\end{equation}
and factor its bispinor as
\begin{equation}
k_{\alpha\dot\alpha}=k_\mu\sigma^\mu_{\alpha\dot\alpha}
 =\kappa_\alpha\widetilde\kappa_{\dot\alpha}~.
\end{equation}
Spinor indices are raised and lowered with
$\epsilon^{01}=-\epsilon_{01}=1$, and
\begin{equation}
 \langle ab\rangle=\epsilon_{\alpha\beta}a^\alpha b^\beta,
 \qquad
 [ab]=\epsilon_{\dot\alpha\dot\beta}\widetilde a^{\dot\alpha}
 \widetilde b^{\dot\beta}~.
\end{equation}
We let $n^\mu=(1,0,0,0)$ denote the radial unit vector and use it to identify dotted and
undotted indices:
\begin{equation}
 \bar\kappa_\alpha=n_{\alpha\dot\alpha}\widetilde\kappa^{\dot\alpha},
 \qquad
 \langle\kappa\bar\kappa\rangle=-2i k,
 \qquad
 k^{\alpha\beta}=\kappa^\alpha\bar\kappa^\beta
 =\kappa^{(\alpha}\bar\kappa^{\beta)}+i k\,\epsilon^{\alpha\beta}~.
\end{equation}
The symmetric part is the boundary three-momentum, whereas the antisymmetric part is its
radial component. The radial-gauge helicity polarizations are \cite{Maldacena:2011nz,Armstrong:2020woi}:
\begin{equation}
\label{eq:epspm}
 \epsilon^-_{\alpha\beta}(\vec{k})=\frac{\kappa_\alpha\kappa_\beta}{i k},
 \qquad
 \epsilon^+_{\alpha\beta}(\vec{k})=\frac{\bar\kappa_\alpha\bar\kappa_\beta}{i k}~.
\end{equation}
They are symmetric, null, transverse to $\vec{k}$, and have zero radial component.
For comparison, the usual four-dimensional polarization vectors with reference spinors
$r_\alpha$ and $\widetilde r_{\dot\alpha}$ are
\begin{equation}
\label{eq:epm}
 e^-_{\alpha\dot\alpha}(\vec k)
 =\sqrt{2}\,\frac{\kappa_\alpha\widetilde r_{\dot\alpha}}{[r\kappa]},
 \qquad
 e^+_{\alpha\dot\alpha}(\vec k)
 =\sqrt{2}\,\frac{r_\alpha\widetilde\kappa_{\dot\alpha}}{\langle r\kappa\rangle}~.
\end{equation}
Both vectors
in \eqref{eq:epm} are null and transverse, and a change of reference spinor shifts them by a multiple
of $k_\mu$. We will now see explicitly that \eqref{eq:epm} and \eqref{eq:epspm} are related by a multiple
of $k_\mu$.
\paragraph{Matching the two polarization conventions}
Define $\bar r_\alpha=n_{\alpha\dot\alpha}\widetilde r^{\dot\alpha}$, giving
$[r\kappa]=-\langle\bar r\,\bar\kappa\rangle$. The two-dimensional Schouten identity gives
\begin{equation}
 r_\alpha
 =\frac{\langle r\bar\kappa\rangle}{\langle\kappa\bar\kappa\rangle}\,\kappa_\alpha
 -\frac{\langle r\kappa\rangle}{\langle\kappa\bar\kappa\rangle}\,\bar\kappa_\alpha\,,
 \qquad
 \bar r_\alpha
 =\frac{\langle\bar r\bar\kappa\rangle}{\langle\kappa\bar\kappa\rangle}\,\kappa_\alpha
 -\frac{\langle\bar r\kappa\rangle}{\langle\kappa\bar\kappa\rangle}\,\bar\kappa_\alpha~.
\end{equation}
Substitution into \eqref{eq:epm}, together with $\langle\kappa\bar\kappa\rangle=-2i k$, yields
\begin{align}
\label{eq:eepsrelation1}
 \sqrt{2}\,e^-_{\alpha\dot\alpha}(\vec k)n_\beta{}^{\dot\alpha}
 =\frac{\kappa_\alpha\kappa_\beta}{i k}
 -\frac{2\langle\bar r\kappa\rangle}
 {[r\kappa]\langle\kappa\bar\kappa\rangle}
 \kappa_\alpha\bar\kappa_\beta\,,\qquad \sqrt{2}\,e^+_{\alpha\dot\alpha}(\vec k)n_\beta{}^{\dot\alpha}
 =\frac{\bar\kappa_\alpha\bar\kappa_\beta}{i k}
 +\frac{2\langle r\bar\kappa\rangle}
 {\langle r\kappa\rangle\langle\kappa\bar\kappa\rangle}
 \kappa_\alpha\bar\kappa_\beta~.
\end{align}
The factor of $\sqrt{2}$ is present because for the polarization vector we use the rescaled bispinor
$e_{\alpha\beta}=\sqrt{2}\,e_{\alpha\dot\alpha}n_\beta{}^{\dot\alpha}$.
The first term on each right-hand side of \eqref{eq:eepsrelation1} is proportional to \eqref{eq:epspm}, while the second is proportional to
$k_{\alpha\beta}=\kappa_\alpha\bar\kappa_\beta$ and is therefore pure gauge. The unique gauge
transformation that sets the radial component to zero is
\begin{equation}
\epsilon^\pm_\mu(\vec{k})
 =e^\pm_\mu(\vec k)-\frac{e^\pm_0(\vec k)}{k^0}\,k_\mu
 =e^\pm_\mu(\vec k)-\frac{e^\pm_0(\vec k)}{i k}\,k_\mu~.
\end{equation}
Indeed, the zero component of the right-hand side is
$e^\pm_0-e^\pm_0(i k)/(i k)=0$.

\section{Flat-space limit at four points}
This appendix contains the derivations used to obtain the flat-space limit in Section~\ref{sec:Single-minus gluon amplitudes}.

\subsection{General kinematics} \label{app:flat_limit_general}
The three-dimensional Schouten identity reads
\be
\label{eq:3dSchouten}
\epsilon_i^-(\vec k_1)[2^+3^+4^+]-\epsilon_i^+(\vec k_2)[3^+4^+1^-]+\epsilon_i^+(\vec k_3)[4^+1^-2^+]-\epsilon_i^+(k_4)[1^-2^+3^+]=0~.
\ee
Equation~\eqref{eq:3dSchouten} can be contracted with either $\epsilon^-(\vec k_1)$ or $\vec k_1$ to give the following two equations:
\bea
\label{eq:TotE3}
&0=\epsilon_i^-(\vec k_1)\epsilon_i^+(\vec k_2)[3^+4^+1^-]-\epsilon_i^-(\vec k_1)\epsilon_i^+(\vec k_3)[4^+1^-2^+]+\epsilon_i^-(\vec k_1)\epsilon_i^+(\vec k_4)[1^-2^+3^+]\,,\\
&0=k_{1i}\epsilon_i^+(\vec k_2)[3^+4^+1^-]-k_{1i}\epsilon_i^+(\vec k_3)[4^+1^-2^+]+k_{1i}\epsilon_i^+(\vec k_4)[1^-2^+3^+]~.
\eea
The second equation can be rewritten as
\be
\label{eq:TotE2}
0=-\frac{\epsilon_i^-(\vec k_1)\epsilon_i^+(\vec k_2) k_{12}^\mu k_{12\mu}}{\epsilon^-(\vec k_1)\cdot \vec k_2}[1^-3^+4^+]+\frac{\epsilon_i^-(\vec k_1)\epsilon_i^+(\vec k_4) k_{14}^\mu k_{14\mu}}{\epsilon^-(\vec k_1)\cdot \vec k_4}[1^-3^+2^+]
+\frac{\epsilon_i^-(\vec k_1)\epsilon_i^+(\vec k_3) k_{13}^\mu k_{13\mu}}{\epsilon^-(\vec k_1)\cdot \vec k_3}[1^-2^+4^+]~.
\ee
Four-momentum conservation implies $k_{12}^\mu k_{12\mu}+k_{13}^\mu k_{13\mu}+k_{14}^\mu k_{14\mu}=0$ and $\epsilon^-(\vec k_1)\cdot \vec k_2+\epsilon^-(\vec k_1)\cdot \vec k_3+\epsilon^-(\vec k_1)\cdot \vec k_4=0$, so the last term can be rewritten as
\be
\frac{ k_{12}^\mu k_{12\mu}+ k_{14}^\mu k_{14\mu}}{\epsilon^-(\vec k_1)\cdot \vec k_2+\epsilon^-(\vec k_1)\cdot \vec k_4} \epsilon_i^-(\vec k_1)\epsilon_i^+(\vec k_3)[1^-2^+4^+]~.
\ee
Using the first equation in \eqref{eq:TotE3}, we obtain
\begin{equation}
\begin{split}
&\frac{\epsilon_i^-(\vec k_1)\epsilon_i^+(\vec k_3) k_{13}^\mu k_{13\mu}}{\epsilon^-(\vec k_1)\cdot \vec k_3}[1^-2^+4^+]   \\
&=-\frac{ k_{12}^\mu k_{12\mu}+ k_{14}^\mu k_{14\mu}}{\epsilon^-(\vec k_1)\cdot \vec k_2+\epsilon^-(\vec k_1)\cdot \vec k_4} (-\epsilon_i^-(\vec k_1)\epsilon_i^+(\vec k_2)[1^-3^+4^+]+\epsilon_i^-(\vec k_1)\epsilon_i^+(\vec k_4)[1^-3^+2^+])~.
\end{split}
\end{equation}
Substituting this back into \eqref{eq:TotE2} yields \eqref{eq:magicCancel}.

\subsection{Half-collinear locus} \label{app:flat_limit_collinear}

The three-dimensional polarization vectors $\epsilon_i^\pm(\vec{k}_i)$ can be extended to four-dimensional polarization vectors $\epsilon_\mu^\pm(\vec{k}_i)$ by setting $\epsilon_0^\pm(\vec{k}_i)=0$. This is precisely our radial-gauge condition \eqref{radial gauge}.
As shown in Appendix~\ref{App:spinorHelicity}, the relation between these polarization vectors $\epsilon_\mu^\pm(\vec{k}_i)$ and the standard four-dimensional polarization vectors $e_\mu^\pm(\vec k_i)$ used in \cite{Guevara:2026qzd} is
\be
\label{eq:DifferentPolarizations}
 \epsilon_\mu^\pm(\vec{k}_i)= e_\mu^\pm(\vec k_i)-\frac{e_0^\pm(\vec k_i)}{ik_i}k_{i,\mu}~.
\ee
Using \eqref{eq:DifferentPolarizations}, we find
\be
 \epsilon^+(\vec k_2)\cdot \vec k_1=  \epsilon^+_\mu(\vec k_2) k_1^\mu=e^+_\mu(\vec k_2) k_1^\mu-\frac{e_0^+(\vec k_2)}{ik_2}k_{1\mu}k_2^\mu~,
\ee
where the second term is proportional to $2k_{1\mu}k_2^\mu = k_{12\mu}k_{12}^\mu$ and is therefore regular on the half-collinear locus. It contributes only to the regular part, which we previously showed to vanish. However, the first term is
\be
\frac{2i e_\mu^+(\vec k_2) k_1^\mu} {k_{12}^\mu k_{12\mu}+i \epsilon}=\sqrt{2}i \frac{\langle r1\rangle}{\langle r2\rangle} \frac{[12]} {\langle 12\rangle[12]+i \epsilon}= \sqrt{2}i \frac{\langle r1\rangle}{\langle r2\rangle} \text{PV}\frac{1} {\langle 12\rangle}+\frac{1}{\sqrt{2}}\,\text{sg}_{12}\frac{\langle r1\rangle}{\langle r2\rangle} \delta(\langle 12\rangle)~,
\ee
where we use the normalization $\int dx\, \delta(x)=2\pi$ of \cite{Guevara:2026qzd}. Again, the principal value on the right-hand side contributes to the regular part, which we showed above to vanish, so we need only consider the second term. It includes another delta function that combines with four-momentum conservation to give
\be
\delta(\langle12\rangle )\delta^3\bigg(\sum_{i=1}^4 \vec{k}_i\bigg)\delta(E)\propto \delta(\langle12\rangle ) \delta^4\bigg(\sum_{i=1}^4 k_i^\mu\bigg)~,
\ee
which can be rewritten in the following way.
First, choose $|r\rangle$ such that $\langle r1\rangle\neq 0$. Any external undotted spinor can be rewritten in the basis $(|1\rangle,|r\rangle)$ as
\be
|i\rangle=\frac{\langle ri\rangle}{\langle r1\rangle}|1\rangle-\frac{\langle 1i\rangle}{\langle r1\rangle}|r\rangle~,
\ee
so that 
\be
\sum_{i=1}^4 (k_i)_{\alpha \dot{\alpha}}= \frac{1}{\langle r1\rangle}\bigg(|1\rangle \sum_{i=1}^4 \langle ri\rangle[i|- |r\rangle \sum_{i=1}^4 \langle 1i\rangle[i|\bigg)~.
\ee
Using $\delta^4(A\cdot x)=|\det (A)|^{-1}\delta^4(x)$, this gives
\be
\delta^4\bigg(\sum_{i=1}^4 k_i^\mu\bigg)=4\langle r1\rangle^2 \delta^2\bigg(\sum_{i=1}^4  \langle ri\rangle[i|\bigg)\delta^2\bigg(\sum_{i=1}^4  \langle 1i\rangle[i|\bigg)~.
\ee
Assuming $[34]\neq 0$, we can again use $\delta^2(A\cdot x)=|\det (A)|^{-1}\delta^2(x)$ to obtain
\begin{equation}
\begin{split}
\delta(\langle12\rangle ) \delta^2\bigg(\sum \langle 1i\rangle \tilde{\lambda}_i \bigg)&=\delta(\langle12\rangle ) \delta^2\bigg( \langle 12\rangle[2|+\langle 13\rangle[3|+\langle 14\rangle[4| \bigg)\\
&=\delta(\langle12\rangle ) \delta^2\bigg(\langle 13\rangle[3|+\langle 14\rangle[4| \bigg)\\
&=\frac{1}{|[34]|}\delta(\langle12\rangle ) \delta(\langle 13\rangle) \delta(\langle 14\rangle)
~.
\end{split}
\end{equation}
This means
\be
\delta(\langle12\rangle ) \delta^4\bigg(\sum_{i=1}^4 k_i^\mu\bigg)=
\frac{ 4\langle r1\rangle^2 }{|[34]|}\delta(\langle12\rangle ) \delta(\langle 13\rangle) \delta(\langle 14\rangle)\delta^2\bigg(\sum_{i=1}^4  \langle ri\rangle[i|\bigg)~.
\ee
The remaining quantity to compute is $k_1[134]$. Because all $\epsilon^\pm_0(\vec k_i)=0$, we can rewrite it using the four-dimensional Levi--Civita symbol $\epsilon_{\mu \nu\rho\sigma}$:
\begin{equation}
\begin{split}
k_1[1^-3^+4^+]=&-i\epsilon^{\mu \nu\rho\sigma} k_{1\mu} \epsilon^-_\nu(\vec k_1)\epsilon^+_\rho(\vec k_3)\epsilon^+_\sigma(\vec k_4)\\
=&
-i\epsilon( k_{1}, e^-(\vec k_1),e^+(\vec k_3),e^+(\vec k_4))+ \frac{e^+_0(\vec k_3)}{k_3}\epsilon( k_{1}, e^-(\vec k_1),k_3,e^+(\vec k_4))\\
&+ \frac{e^+_0(\vec k_4)}{k_4}\epsilon( k_{1}, e^-(\vec k_1),e^+(\vec k_3),k_4)+i\frac{e^+_0(\vec k_3)e^+_0(\vec k_4)}{k_3k_4}\epsilon( k_{1}, e^-(\vec k_1),k_3,k_4)
~.
\end{split}
\end{equation}
We then use
\begin{equation}
\begin{split}
&-4\epsilon(|A_1\rangle[B_1|,|A_2\rangle[B_2|,|A_3\rangle[B_3|,|A_4\rangle[B_4|) \\
&=\langle A_1A_2\rangle[B_2B_3]\langle A_3A_4\rangle[B_4B_1]-[B_1B_2]\langle A_2A_3\rangle[B_3B_4]\langle A_4A_1\rangle~,
\end{split}
\end{equation}
to see that the first term alone is not proportional to $\langle 13\rangle$ or $\langle 14\rangle$, which gives
\be
k_1[1^-3^+4^+]=-\frac{i}{\sqrt{2}}[34]\frac{\langle r1\rangle^2}{\langle r3\rangle\langle r4\rangle} \qquad \text{ modulo } \langle 13\rangle \text{ and } \langle 14\rangle~.
\ee
Putting all this together implies
\begin{equation}
\begin{split}
&k_1[1^-3^+4^+]\text{sg}_{12}\frac{\langle r1\rangle}{\langle r2\rangle} \delta(\langle 12\rangle) \delta^3\bigg(\sum_{i=1}^4 \vec{k}_i\bigg)\delta(E)\\
&\propto -\text{sg}_{12}\text{sg}_{34} \frac{\langle r1\rangle^5}{\langle r2\rangle\langle r3\rangle\langle r4\rangle} \delta(\langle 12\rangle) \delta(\langle 13\rangle) \delta(\langle 14\rangle) \delta^2\bigg(\sum \langle ri\rangle \tilde{\kappa}_i \bigg)~,
\end{split}
\end{equation}
where we defined 
\begin{equation} \label{eq:sgn}
    \text{sg}_{ij} \equiv \mathrm{sign}([i j ] ) \;.
\end{equation}
This is the $s$-channel contribution to the expected single-minus gluon amplitude \cite{Guevara:2026qzd}. Adding the $t$-channel then leads to \eqref{eq:finalamplitude}.

\addcontentsline{toc}{section}{References}
\bibliographystyle{style}
\bibliography{references}

\end{document}